\documentclass{article}
\usepackage{graphicx} 
\usepackage{xcolor}
\usepackage{biblatex}
\usepackage{tabularx}
\usepackage{subfig}
\usepackage{verbatim}
\usepackage{authblk}

\title{From Airspace Degradation to Deconfliction: Predicting Safety-Critical Traffic States through the Evolution of Multiscale Complexity Metrics}

\author[1, $\ast$]{Ra\'ul L\'opez-Mart\'in}
\author[2]{Justus Renkhoff}
\author[3]{Yash Guleria}
\author[2]{Mohsan Jameel}
\author[4]{Massimiliano Zanin}

\affil[1]{Instituto de F\'isica Interdisciplinar y Sistemas Complejos (IFISC), CSIC-UIB, Campus UIB, 07122 Palma, Spain}

\affil[2]{Institute of Flight Guidance, German Aerospace Center, 38108 Braunschweig, Germany}

\affil[3]{School of Management, Indian Institute of Technology (IIT) Mandi, H.P., 175005, India}

\affil[4]{Instituto de F\'isica Interdisciplinar y Sistemas Complejos (IFISC), CSIC-UIB, Parc Bit, 07120 Palma, Spain}

\affil[$\ast$]{Corresponding author: \texttt{raullopez@ifisc.uib-csic.es}}

\begin{document}

\maketitle

\begin{abstract}
Growing air traffic demand is pushing airspace management towards the limits of its operational capacity, with steadily increasing workload for air traffic control officers, with direct consequences for safety. Understanding which traffic conditions give rise to safety-critical situations is therefore an essential open problem.
Here we compare a large number of deconfliction events with equivalent conflict-free scenarios, extracted from thirteen weeks of air traffic control data over the Swedish Flight Information Region. We describe the traffic structure before each event through thirty sector-complexity metrics computed at multiple spatial and temporal scales, and use them as inputs of Machine Learning models.
We find that the need of deconflictions can be predicted with an accuracy reaching $70\%$, but only when the traffic is described at the sector spatial scale. Predictability increases monotonically in the minutes preceding the clearance, showing that situations requiring deconflicting actions are not sudden occurrences, but rather the endpoint of a progressive degradation process. Additionally, the information needed for such prediction mainly resides in specific characteristics of the overall traffic structure, and not only in the conflicting pair.
These results numerically demonstrate that safety-critical airspace states can be anticipated from the collective behaviour of the traffic, paving the way for early-warning systems that flag degraded, safety-relevant conditions before an intervention is required.
\end{abstract}


\section{Introduction}

The continuously increasing demand on air traffic is pushing the system to the limits of its operational capacity. The negative impact of this is two-fold: it both increases the workload of all the actors involved in the management of air traffic, and also makes such management a more challenging task. Safety is definitely the most sensitive domain to be impacted by this situation, as a deterioration of the former could potentially lead to life-threatening situations; hence, the optimisation of the system, here understood as the reduction of its complexity, is nowadays more crucial than ever.

Manipulating and optimising any system, including air traffic, necessarily hinges on understanding it. In this context and from a safety viewpoint, this requires comprehending under which conditions risk-related (and safety-concerning) scenarios arise, in order to push the system away from those conditions through operations and management.
In the domain of air traffic management, this has traditionally been tackled through small-scale studies, focusing on the activities performed by Air Traffic Control Officers (ATCOs) \cite{zhang2015from}; only more recently, the emergence of new technologies in data analysis has enabled the introduction of machine learning (ML) \cite{gallego2018analysis, pham2020air, gutierrez2022study,  guleria2024towards} and artificial intelligence (AI) \cite{bastas2022data, gerdes2025synergies, kent2026future} solutions.
Finding optimised trajectories \cite{bastas2022data, nebula2023digital}, assisting ATCOs \cite{tyburzy2025user} in their decisions or managing sector capacity \cite{tang2012multi, liu2025multi} are, among others, fields in which these techniques have recently been employed to aid in air traffic operations. 
Important success stories notwithstanding, these two families of approaches also present important limitations. Specifically, the former ones can provide detailed insights into the actuation of ATCOs, but are usually limited in scope, due to the difficulty of performing large-scale experiments involving professional participants. On the other hand, the latter ones can provide excellent solutions and recommendations; yet, their black-box nature hinders the extraction of actionable knowledge.

This latter aspect, i.e. obtaining actionable insights about how the system works and under which conditions safety may be degraded, has important consequences in many operational and research contexts.
To illustrate, a better knowledge of ATCO decision-making processes, and their natural inter-personal variability, can support a more precise monitoring and management of their workload; in other words, situations too complex to be safely handled can be forecasted and avoided before they completely develop. 
This can then be used to improve existing software tools, e.g. real-time decision support tools and early warning systems \cite{tyburzy2025user}, or for strategic and tactical dynamic airspace sectorisation \cite{gerdes2018dynamic, kopardekar2007initial}. Critically, knowing how ATCO take decisions can allow to adapt these tools to the human way of working, instead of enforcing the synthetic one to the operators \cite{inoue2015applying, timotic2022automation}.
Finally, being able to understand airspace conditions that are associated with risk-related events, or conversely with safe ones, can further pave the way to the creation of more realistic synthetic conflict scenarios, which can then be used to improve training - a possibility that is being explored in other aspects of air transport \cite{esteve2026generation, xiao2016atcem}. 

We here present a data-driven analysis of the airspace conditions that have the largest influence in defining how a traffic situation will evolve, specifically focusing on whether a deconfliction directive has to be given in order to avoid a Loss of Separation (LoS). This is done by extracting deconfliction situations from multiple weeks of operations over the Swedish Flight Information Region (FIR); and equivalent events that did not require any ATCOs intervention.
We next describe the airspace status prior to the event, through classical sector complexity metrics calculated at different spatial and temporal scales. 
These metrics are finally used as inputs of ML models, trained to forecast whether a deconfliction action will take place. These allow us to understand how the predictability of such actions evolves in space (i.e. as a function of the airspace under analysis) and time (i.e. in terms of the time remaining to the clearance); and to further identify which metrics underpin such predictability.

The main result that will be shown throughout this work is that deconfliction events can be classified, and hence predicted. We find that these events are in general not sudden occurrences, but rather emerge at the end of a dynamical process that spans several minutes; the need of the intervention of ATCOs is thus not instantaneous, but rather something that builds up with time. Factors associated with the prediction include, among others, the traffic density, the distance of aircraft with the airspace borders, and the variability in speed and altitude. Additionally, these factors have to be evaluated at specific spatial scales, as e.g. including data from a whole FIR may dilute the useful information.
In synthesis, these results indicate that it is in principle possible to create a warning system, alerting the ATCOs when a complex situation outside their control is developing; and further provide indications on the characteristics of the traffic that ought to be monitored. 
Consequently, the main contributions of this work include the first comparative analysis of deconfliction vs. conflict-free situations using machine learning tools; the demonstration that deconflictions are predictable and build up over time; the identification of the optimal temporal and spatial scales in which such predictability is more easily accessible; and the identification of the metrics underpinning such predictability.

In the remainder of this work we first introduce, in Sec.~\ref{sec:sota}, a review of the state of the art on evaluating the conditions leading to the appearance of safety-related events. Next, Sec.~\ref{sec:methods} introduces the data and methods buttressing this work. Specifically, Sec.~\ref{sec:operational_data} presents the catalogue of deconfliction events; Secs.~\ref{sec:complexity_metrics} and \ref{sec:spatial_and_temporal_ranges} describe the sector complexity metrics and their spatial and temporal scales; and Secs. \ref{sec:class_prob_def} and \ref{sec:class_mods_def_comp_and_tun} frame the analysis as a machine learning classification problem.
The results of the proposed analyses are presented in Sec.~\ref{sec:results}, including the impact and selection of the temporal and spatial scales (Sec.~\ref{sec:spat_and_temp_effect}); the identification of the metrics are the most meaningful to distinguish deconflict events (Sec.~\ref{sec:metrics_distributions_comparisons}); and the characterisation of their temporal evolution (Sec.~\ref{sec:time_evol_anal}).
To conclude, Sec.~\ref{sec:discusion} will discuss the main findings, their operational implications, limitations, and future research directions.

\section{State of the art}
\label{sec:sota}

Understanding what conditions, associated to all facets of a transport system, lead to the appearance of safety-related events is of the utmost importance; as only by securing this knowledge can safe and efficient operations be guaranteed.
The Air Traffic Management (ATM) research community has long understood the relevance of this task, and works on this topic can be found as early as in the 1970s \cite{dunlay1975analytical}. Efforts have included a wide variety of different approaches and methodologies regarding, e.g., the type of events under analysis, the way of characterising them, or the considered data source. This section briefly explains these approaches, mentions their main common limitations, and highlights how this work tackles these limitations and further contributes to this field of knowledge.

\subsection{Main characteristics of past approaches}
\label{sec:char_past_appr}

\subsubsection*{Type of events analysed} 

This refers to which events are considered as  safety-related in the analysis. The most common choice in the literature are Losses of Separation (LoS) \cite{valdes2019development, serrano2022identification, garcia2019influence}. These events are situations in which the distance between two aircraft falls below a predetermined distance threshold ($5$ NM horizontally and $1,000$ ft vertically, in the case of the previous works), as such reduced separation is considered a safety-threatening scenario. 
Those thresholds are nevertheless not universal. To illustrate, in Ref.~\cite{perez-castan2022machine} authors decided to use an extended $10$ NM horizontal threshold; while Ref.~\cite{brooker2005reducing} worked with the events known as \textit{Airprox}, i.e. situations in which a professional has considered safety to be compromised due to the distance, relative positions and speeds between the aircraft \cite{icao2016procedures}. These choices represent proxies of situations in which the management of the traffic has gone explicitly wrong, as safety standards have been violated. Conversely, other works choose to focus on situations that would have potentially lead to unsafe scenarios, had the ATCOs not acted to avoid them; these events are known as deconflictions \cite{gaume2025extracting}. Once these events of interest are fixed, the researcher has to further decide how to analyse them. The most direct option is to consider those events individually \cite{brooker2005reducing, gaume2025extracting}. Alternatively, one may classify (or cluster) the events as a function of their type or frequency, in order to highlight differences between subgroups; examples include Ref.~\cite{serrano2022identification} and Refs.~\cite{valdes2019development, garcia2019influence}, respectively. Finally, they can be compared with additional events representing the opposite scenario, i.e. safe ones, with the goal of differentiating between both type of situations, namely unsafe and safe \cite{perez-castan2022machine}.

\subsubsection*{Methodology to characterise each event}

This refers to which property of the scenarios are described. Taking into account that all the types of events mentioned in the previous paragraph involve relations between aircraft on a pairwise fashion, the first big distinction that can be found is the scale of the metrics. Some works characterise the scenario by exclusively focusing on the pair of aircraft generating the event, hence neglecting any possible surrounding traffic \cite{gaume2025extracting, perez-castan2022machine, brooker2005reducing, serrano2022identification}; others evaluate the overall situation of the airspace \cite{valdes2019development, garcia2019influence}. For the latter ones, different points of view are considered; to illustrate, Ref.~\cite{valdes2019development} focuses on the traffic flows, while Ref.~\cite{garcia2019influence} pays special attention to the situation around waypoints.

\subsubsection*{Data sources} 

The results of any data-driven study can only be as good as the underlying data; hence the availability of data sources and their quality are pivotal elements in the field. Focusing first on the works that use LoSs events (or related concepts), as these events are defined from a pairwise distance relationship between aircraft, one approach is to extract them from executed trajectories \cite{valdes2019development, garcia2019influence}. This approach has also been extended by modifying the trajectories, synthetically creating (and hence, increasing the number of) LoS events available for analysis \cite{perez-castan2022machine}. An alternative is to rely on historical records of LoS \cite{serrano2022identification} or Airprox \cite{brooker2005reducing} events. Finally, the identification of deconfliction events from executed trajectories has been augmented by comparing them with planned ones \cite{gaume2025extracting}.

\subsection{Common limitations}

\subsubsection*{Information to characterise the events}

As mentioned in the second paragraph of Sec.~\ref{sec:char_past_appr}, some works describe the events of interest based on the properties of the two aircraft participating in them \cite{gaume2025extracting, perez-castan2022machine, serrano2022identification}. While those aircraft are indeed of high relevance, disregarding the surrounding traffic can lead to a misrepresentation of the global situation, or to an incomplete picture of it. On the other hand, works that do consider the rest of the traffic still present limitations. To illustrate, Ref.~\cite{garcia2019influence}, by focusing exclusively on the situation around waypoints, does not capture the information associated to entire sectors, which is the natural area of responsibility of ATCOs. Conversely, Ref.~\cite{valdes2019development} describes the status through the characteristics of the traffic flows over large regions of the airspace; this entails two limitations. First, the grouping of aircraft in traffic flows, while being a way to simplify the analysis, also simplifies the airspace itself. Additionally, and most significantly, the fact that these flows are studied in large volumes of airspace, broader than the areas of responsibility of ATCOs, implies that more information, i.e. more than what the ATCO has, is being added to the characterisation of the events - as the work itself acknowledges \cite{valdes2019development}. Finally, studies like Ref.~\cite{brooker2005reducing}, which rely on more qualitative descriptions of the situations, lack generalisability.

\subsubsection*{Events considered for the analysis} 

Two major limitations emerge from the events themselves. The first one is related to the their definition and identification. On the one hand, works that identify LoSs events through trajectory analysis present severe limitations. To illustrate, Ref.~\cite{perez-castan2022machine} generates synthetic events by modifying the times of entrance of aircraft in the sector; while more instances are available, these are not real, being this a major drawback from the viewpoint of the trustfulness of the results. Refs.~\cite{garcia2019influence, valdes2019development} do use only real events extracted from executed trajectories; these are nevertheless limited by the simplicity of the extraction methodology, and by the lack of complementary data for validation. 
On the other hand, one can work exclusively with LoSs (or equivalent events) registered by aviation authorities \cite{serrano2022identification, brooker2005reducing}. While these events have excellent quality and reliability, the number of them rapidly fall to less than $100$, a range in which statistically significant and generalisable results are hard to obtain. 
The second limitation emerges from how events, after being identified, are classified. Some works, namely Refs.~\cite{valdes2019development, garcia2019influence}, correlate the airspace status with the accumulated number of LoSs events, grouping these events, hence moving away from an individualised analysis. This granularity can lead to loosing relevant information from them. Other works, such as Ref.~\cite{gaume2025extracting}, only characterise one type of events (in this case, deconfliction events); yet, the lack of comparison with other types of events does not allow to distinguish whether a characteristic is actually relevant, or is rather a common feature of the airspace or the traffic.

\subsection{Contributions of this work}

The present study builds on the previously reviewed papers and leverages the lessons from them learnt, to yield a set of novel contributions to the literature:

\begin{itemize}
    \item {\it First work comparing deconfliction events with conflict-free scenarios:} This contrastive analysis allows to focus on the differences between the two, and hence to highlight the antecedents of deconfliction actions, by also avoiding confounding properties of the airspace and traffic structure.

    \item {\it Robust detection of deconfliction events, integrating information from ATCO clearances and aircraft maneuvers:} These are identified both through ATCOs' commands included in the dataset \cite{nilsson2023swedish}, and through altitude and heading changes. Compared to the LoS events usually considered in the literature, deconflictions are more common occurrences, thus supporting more robust statistical analyses; at the same time, they still retain a clear link with the safety of the system.

    \item {\it Statistical validation of airspace and traffic features:} The use of ML models allows to assess the relevance of specific properties of the traffic patterns in an objective way. It further allows to compare a large number of properties, $30$ in the present case, going well beyond similar studies in the literature.

    \item {\it Analysis of the optimal spatial and temporal scales of the analysis:} Compared to other studies, here both the spatial and temporal scales for describing events are optimised using ML models; they are therefore data-driven and hypotheses-free. This maximises the quantity of information about neighbouring traffic that is included in the analysis, while minimising the inclusion of confounding effects.

    \item {\it Demonstration of the progressive degradation of safety:} We present numerical evidences suggesting that a deconfliction is the result of a progressive degradation of the safety as a consequence of evolving traffic patterns, over periods of up to ten minutes, as opposed to instantaneous occurrences.
\end{itemize}

\section{Methods}
\label{sec:methods}

\subsection{Operational data}
\label{sec:operational_data}

The data used in this work come from the Swedish Civil Air Traffic Control (SCAT) dataset \cite{nilsson2023swedish}, which includes 13 weeks of air traffic control data for the Swedish airspace. The dataset contains detailed records for nearly $170,000$ scheduled flights, including flight plans, ATCOs issued clearances, weather information, and airspace information. Additional characteristics are reported in Tab. \ref{tab:data}.

\begin{table}[!tb]
\centering

\renewcommand{\arraystretch}{1.2}
\setlength{\tabcolsep}{5pt}
\begin{tabularx}{\linewidth}{l c}
\hline
    Temporal range & From 2016-10-15 to 2017-09-22 \\ \hline
    Spatial range & SWEDEN FIR (ESAA)\\ \hline
    Number of flights & 167,547 \\ \hline
    Average temporal resolution & 5 secs between reports \\ \hline
    Number of deconflictions & 646 \\ \hline
    Of which vertical / horizontal & 518 / 128 \\ \hline
    Total data size & 51.1 GB \\ \hline
\end{tabularx}
\caption{Characteristics of the Swedish Civil Air Traffic Control (SCAT) dataset \cite{nilsson2023swedish}, and details about the deconfliction events from it extracted.}
\label{tab:data}
\end{table}

\begin{figure}[!tb]
\centering
\includegraphics[width=1\columnwidth]{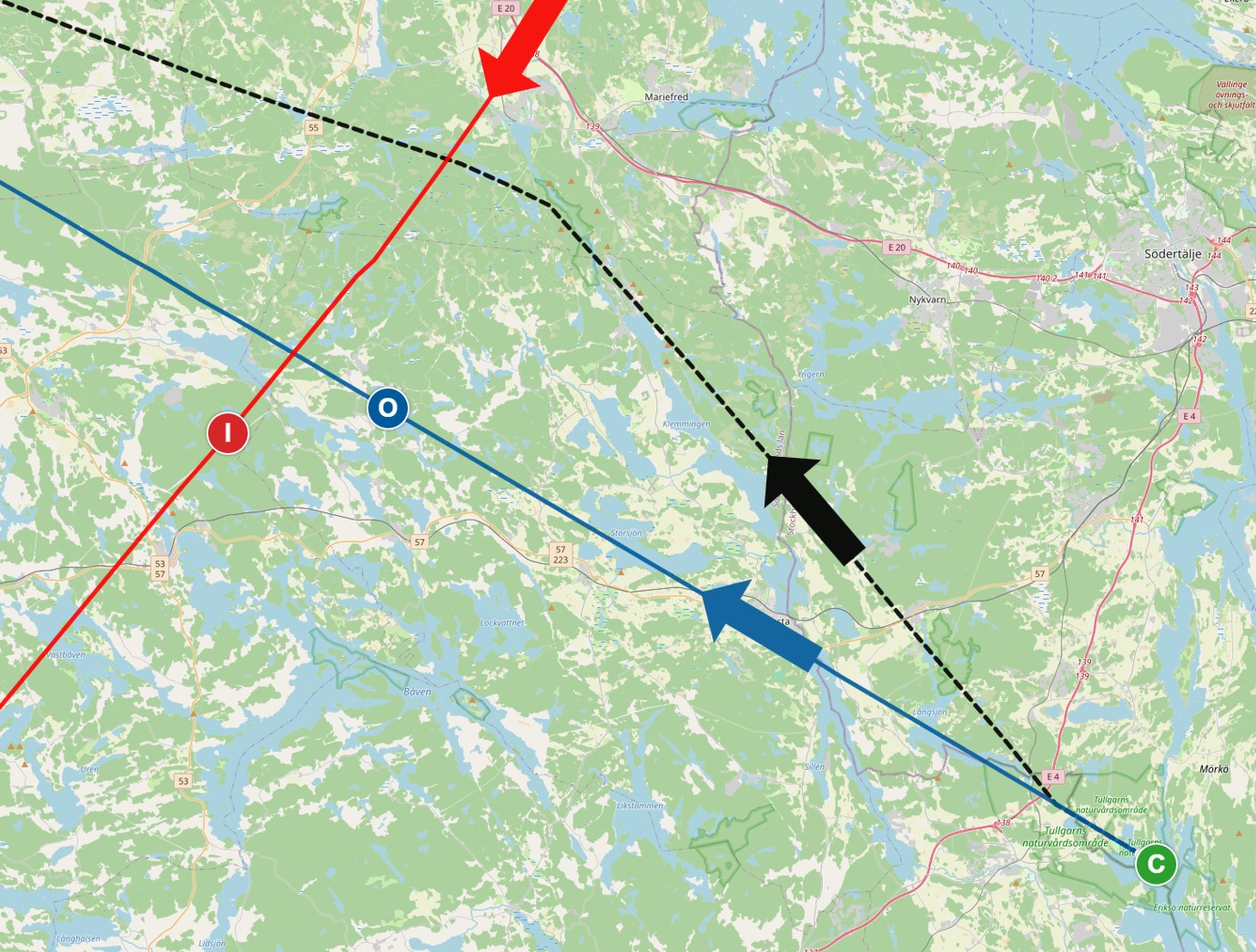}
\caption{Example of a potential conflict from the SCAT dataset that was resolved by an ATCO-issued heading clearance. The lines represent trajectories, including the actual flight path of the ownship (black, dashed), its projected trajectory without clearance (blue, solid), and the intruder's flight path (red, solid). The green marker indicates the location at which a clearance was issued, while the red and blue ones the positions at which the conflict between the aircraft would have occurred. Map provided by OpenStreetMap \cite{openstreetmap}.}
\label{fig:conflict}
\end{figure}

The information contained allows the detection of potential conflicts that were resolved by ATCOs \cite{renkhoff2025identifying}. Specifically, we identify heading and flight level manoeuvrers that can be linked to prior ATCOs clearances, and test whether these interventions prevented a potential conflict. For each relevant clearance, a projected no-clearance trajectory is reconstructed. In the case of heading changes, this is done by assuming a direct course towards the next waypoint according to the flight plan; conversely, for flight level changes, the horizontal component of the flown trajectory is preserved, while the altitude is kept constant at the pre-clearance one. These projected trajectories are then checked against surrounding traffic using predefined spatial and temporal thresholds, to identify separation minima infringements. If the actual trajectory provides greater separation than the projected no-clearance trajectory, the clearance is interpreted as a deconfliction \cite{renkhoff2025identifying}. In the former case, a conflict time is further defined, corresponding to the timestamp in which the projected trajectories reach the smallest lateral separation. A graphical representation of a potential conflict that was resolved using a heading clearance is reported in Fig. \ref{fig:conflict}.

Regarding the identification of an event as a potential LoS, it is worth noting that different thresholds from the legal ones (5\,NM horizontally and 1000\,ft vertically) have here been used. A vertical threshold of 900\,ft was selected to reduce false positives, as small deviations below 1000\,ft may result from sensor inaccuracies rather than reflecting an actual loss of separation. In contrast, the horizontal threshold was increased to 8\,NM. This choice was motivated by findings from recent research works showing that ATCOs often maintain larger horizontal margins in practice, frequently around 8\,NM \cite{gaume2025extracting}. The most important parameters used in the identification of LoSs are reported in Tab. \ref{tab:conflict_params}.

\begin{table}[!tb]
\centering
\renewcommand{\arraystretch}{1.2}
\setlength{\tabcolsep}{5pt}
\begin{tabularx}{\linewidth}{l c X}
\hline
\textbf{Parameter} & \textbf{Value} & \textbf{Description} \\
\hline
Horizontal separation threshold & 8\,NM & Horizontal distance threshold below which two aircraft are considered to be in potential conflict \\
Vertical separation threshold & 900\,ft & Vertical separation threshold below which two aircraft are considered to be in potential conflict \\
Detection horizon & 1200\,s & Look-ahead time after the clearance within which projected conflicts are searched \\
Ownship projection interval & 5\,s & Temporal resolution used to sample the projected ownship trajectory along the next flight-plan waypoints \\
Temporal matching tolerance & $\pm$5\,s & Time window around each projected ownship position in which intruder traffic samples are considered \\
\hline
\end{tabularx}
\caption{Parameters used for defining the presence of a conflict in the pre-clearance trajectories.}
\label{tab:conflict_params}
\end{table}

\subsection{Complexity metrics}
\label{sec:complexity_metrics}

Within the available alternatives for characterising and measuring the difficulty of managing the traffic within a sector, and hence for quantifying ATCOs workload, sector complexity metrics stand out as a valuable option. These are quantitative measures, calculated on the traffic within a sector of the airspace, that numerically describe a specific property of it, and that are assumed to be related to the complexity of its management. Literature on this topic can be traced back to the 1970s \cite{hurst1978objective}; and by the early 2000s, the number of proposed sector complexity metrics was already over one hundred \cite{hilburnl2004cognitive}. This trend has not slowed down \cite{moreno2022determination, prandini2011toward}, and this topic is still an open and active field of research \cite{lui2026temporal, lopez2026towards}. 
Given the objective of this work, complexity metrics were identified as the most natural approach for describing air traffic patterns and their future evolution. 
Following a review of the extant literature, a large set of commonly used sector complexity metrics has been identified, implemented, and numerically calculated on each clearance event. These are listed in Tab. \ref{tab:compl_metrics}, grouped by families and with synthetic descriptions; the interested reader can find extended descriptions, implementation details and remarks in App.~\ref{app:compl_metrics}.

\begin{table}[!tb]
    \centering
    \resizebox{\linewidth}{!}{
    \begin{tabular}{l|l|c|c|c|c}
         {\bf Group} & {\bf Metric} & {\bf Refs.} & {\bf Min.} & {\bf Median} & {\bf Max.} \\\hline
         Distance-based & Mean horizontal distance & \cite{chatterji2008analysis} & $1.4$ NM & $51.8$ NM & $225.6$ NM \\ 
          & Mean vertical distance & \cite{chatterji2008analysis} & $0$ FL & $20$ FL& $182$ FL \\
          & Pairwise horizontal  & \cite{stein1985workload, histon2002structural} & $0.0$ & $1.56$ & $3.19$ \\
                    & distances entropy &  &  &  & \\
          & Number of aircraft close to sector border & \cite{prandini2011toward} & $0$ & $2$ & $12$ \\
          & Mean traffic convergence/divergence & \cite{delahaye2000air, xiao2016atcem}  & $-1.492$ NM/ $5$ s & $-0.025$ NM/ $5$ s &  $1.461$ NM/ $5$ s \\
          & Number of pairs with reduced separation &  \cite{eurocontrol2000complexity} & $0$& $0$ & $6$ \\\hline
    Position-based & Number of different flight levels in use &  \cite{moreno2022determination} & $1$  & $3$ & $10$ \\ 
     & Flight levels occupancy entropy  & \cite{majumdar2002factors} & $0.00$ & $1.04$ & $2.25$ \\
     & Heading angle coherence &  \cite{brinton1996dynamic} & $0.002$ & $0.494$ & $1.000$\\\hline
     Speed-based & Mean ground speed & \cite{xiao2016atcem} & $284$ kt& $447$ kt& $559$ kt \\
      & Ground speeds variance & \cite{xiao2016atcem} & $0.01$ kt& $21.50$ kt& $123.33$ kt\\
      & Mean vertical speed & \cite{gianazza2006selection} & $0.00$ FL / s & $0.02$ FL / s& $0.55$ FL / s\\\hline
      Crossings-based & Number of potential crossings &  \cite{brinton1996dynamic} & $0$ & $0$ & $8$ \\
      & Number of potential crossings with & \cite{gianazza2006selection} &  $0$& $0$ & $8$\\
      & convergence angle $> 20^{\circ}$ & & & &  \\ 
      & Number of potential conflicts &  \cite{laudeman1998dynamic} & $0$ & $0$ & $3$ \\
            & Number of potential conflicts with & \cite{laudeman1998dynamic} \cite{gianazza2006selection} & $0$ & $0$ & $2$ \\
      & convergence angle $> 20^{\circ}$ & & & & \\ 
      \hline
      
     Evolution-based & Number of aircraft with absolute &   \cite{laudeman1998dynamic} &  $0$ & $0$ & $7$\\
     & ground speed change $> 10$ kt in 2 min. &  & & &\\
     & Number of aircraft turning &  \cite{laudeman1998dynamic} & $0$ & $0$ & $2$ \\
     & Number of aircraft climbing &  \cite{cardosi1995human} & $0$& $0$ & $7$ \\
     & Number of aircraft accelerating &  \cite{laudeman1998dynamic} & $0$ &  $2$ & $12$\\\hline

     Occupation-based & Accumulated sector occupancy & \cite{hurst1978objective} & $2$ & $16$& $43$\\
     & Past accumulated sector occupancy &  \cite{hurst1978objective} & $2$ & $10$ & $29$ \\
      & Occupancy variation & \cite{chaboud2000air} & $0.12$ & $0.39$ & $1.81$ \\
      & Maximum simultaneous occupation & \cite{hurst1978objective} & $2$& $7$ & $17$\\
      & Mean time flown in sector & \cite{schaefer2001air, majumdar2002factors} & $145$ s & $772$ s & $1594$ s \\
      & Mean distance flown in sector & \cite{schaefer2001air, majumdar2002factors} & $12.6$ NM & $92.3$ NM & $178.5$ NM\\ 
      & Number of entering aircraft & \cite{majumdar2002factors} & $0$& $2$ & $8$ \\
      & Number of exiting aircraft & \cite{majumdar2002factors} & $0$& $2$ & $9$ \\
      & Number of aircraft& \cite{brinton1996dynamic, masalonis2003dynamic} & $2$& $4$& $13$\\ \hline
      Sector-based & Sector area &  \cite{gianazza2006selection}  & $2,630$ NM$^2$ & $10,652$ NM$^2$& $28,201$ NM$^2$\\
      & Sector ID & - & - & - & - \\
      \hline
        
    \end{tabular}
    }
    \caption{List of the considered sector complexity metrics. The first three columns report the general group to which they belong; a synthetic description; and relevant references, including where they were originally defined. The last three columns report the minimum, median and maximum of the observed values (for sectors, and evaluated at clearance time). See App. \ref{app:compl_metrics} for detailed descriptions.}
    \label{tab:compl_metrics}

\end{table}

\subsection{Spatial and temporal scales}
\label{sec:spatial_and_temporal_ranges}

The definition of the traffic snapshot on which complexity metrics are calculated is controlled by two scales, respectively temporal and spatial.

With temporal scale we refer to the time difference (or offset) between the clearance time of the event under study, and the moment when the airspace structure is evaluated - i.e. when complexity metrics are calculated. As the objective of this study is to describe which elements may help forecasting the need for a deconfliction, the considered time offsets will be positive, that is, we analyse the structure of the traffic before the event takes place. 

Moving to the spatial scale, this refers to the portion of the airspace in which the deconfliction event takes place, and on which the complexity metrics are calculated.
Multiple spatial scales can naturally be defined, thanks to both the hierarchical structure of the airspace and the rich information included in the raw dataset \cite{nilsson2023swedish}. We here specifically consider the following ones:

\begin{itemize}
    \item {\it Elementary blocks}. Indivisible volumes of airspace that can be operated independently \cite{ruano2026structural}. They correspond to the smallest airspace subdivision available in the raw dataset \cite{nilsson2023swedish}.

    \item {\it Sectors}. Following the Air Traffic Service (ATS) Airspace Structure Chart \cite{lfv2026sweden} of the Swedish airspace, elementary blocks can be grouped to form sectors, to which the same radio frequency is assigned. The rationale behind the analysis of this division is that areas with common radio frequencies are controlled by the same ATCO (or group thereof); all traffic in them included should therefore affect the decision making process, and in turn, be affected by it.

    \item {\it Area Control Centers (ACCs)}. Individual sectors are merged according to the ACC which they belong to. Two ACCs are defined for the Swedish airspace, namely ESMM (Malm\"o) and ESOS (Stockholm) \cite{nilsson2023swedish}.

    \item {\it Swedish Flight Information Region (FIR)}. The totality of the airspace, corresponding to the complete Swedish FIR. Note that the use of this spatial scale implies that the airspace on which complexity metrics are calculated is the same across all events.
\end{itemize}

\subsection{Definition of the classification problem}
\label{sec:class_prob_def}

As previously introduced, the analyses here reported are going to be based on a classification problem, in which snapshots of the airspace are labelled according to the presence of a deconflicting action. The process to identify both types of snapshots is described below.

\subsubsection{Identification of deconfliction events}
\label{sec:filtering_deconflicting_events}

The first part of the data set is composed of situations in which a deconfliction action has been detected, as described in Sec. \ref{sec:operational_data}. Given the temporal and spatial scales of the analysis, defined in Sec. \ref{sec:spatial_and_temporal_ranges}, only those instances in which two or more aircraft were flying in the evaluated airspace are retained. This is done to ensure that most of the airspace complexity metrics, as previously described in Sec.~\ref{sec:complexity_metrics}, can actually be calculated and yield meaningful values.
Note that this check is independently performed for each temporal and spatial scale. In other words, a snapshot calculated on an elementary block one minute before the deconfliction event may be discarded due to the previous filter; this nevertheless does not preclude the inclusion of another snapshot for the same event, e.g. on the sector scale, or five minutes in advance.

Tab.~\ref{tab:num_deconflictions} presents the number of deconfliction events identified in the data set, for each of the analysed spatial divisions and at clearance time. Due to the aforementioned filter, not all events appear for all the spatial divisions; and, as is to be expected, the more micro the spatial scale is, the lower is the number of events that pass the filter. 

\begin{table}[!tb]
    \centering
    \begin{tabular}{|l|c|c|c|c|}
    \hline
             Spatial division & Elementary blocks & Sectors & ACCs & FIR \\ \hline
             
        Num. of deconflicting events & 341 & 599 & 646 & 646 \\\hline
 
    \end{tabular}
    \caption{Total number of deconflicting events considered in the data set for classification at clearance time, as a function of the used spatial scale.}
    \label{tab:num_deconflictions}
\end{table}

\subsubsection{Identification of conflict-free situations}
\label{sec:conflict_free}

The previous instances are complemented by a same number of snapshots of the system in which no deconflicting manoeuvres were issued, and are therefore labelled as ``conflict-free''. Note that this does not preclude other types of ATCOs interventions and trajectory changes; still, we can assume that no conflict was taking place, nor could potentially have taken place, in that moment in time.
In order to minimise the probability of including confounding effects, these conflict-free situations should be as similar as possible to those including a deconfliction. 
This is achieved by starting from each valid deconfliction event, as described above, and by taking snapshots of the same airspace at the same time, in a different day of the same week. Only those snapshots that satisfy the following two conditions are further retained: they include at least two aircraft, in order to ensure that complexity metrics can be calculated; and they are not associated with any deconfliction action.
Snapshots will thus not include trivial biases, e.g. a deconfliction event in a peak hour shall not be compared with a scenario in the middle of the night. 
As a final point, in order to balance the final data set and counteract the larger number of conflict-free instances that are naturally available in the system, a subset of conflict-free events is randomly drawn in each training/testing iteration.

\subsection{Classification models: definition, comparison and tuning}
\label{sec:class_mods_def_comp_and_tun}

Machine Learning (ML) models stand out as a natural choice to tackle the classification problem described in Sec.~\ref{sec:class_prob_def}, due to their capacity of automatically detecting hidden patterns in large sets of data. Many are available in the literature \cite{hastie2009elements, vapnik2013nature, sayed2023inference}, each with different underlying hypotheses about the patterns that characterise the data, and without simple rules about the selection of the best one for a given problem. We here consequently consider and compare a set of well-established ones, namely: Decision Tree (DT) \cite{maimon2014data, costa2023recent}, Random Forest (RF) \cite{breiman2001random, biau2016random}, Gradient Boosting (GB) \cite{natekin2013gradient, bentejac2021comparative} and Support Vector Machine (SVM) \cite{wang2005support, steinwart2008support}.
In all cases, the \textit{Scikit-learn} implementation has been used \cite{pedregosa2011scikit}. The interested reader can find detailed descriptions of each ML model in App.~\ref{app:class_models}.

Models are compared by assessing their classification performance on the complete balanced data set, i.e. comprising all available deconflicting events and a random subset of the same cardinality of conflict-free ones. Regarding the scales of the analysis, all spatial ones will be individually tested; conversely, from a temporal perspective, the airspace will be described at the moment of the clearance - i.e. for a zero offset, as defined in Sec. \ref{sec:spatial_and_temporal_ranges}. The classification performance will mainly be measured through the accuracy of the model, namely the fraction of events correctly classified; some key results will also be validated through the corresponding F1 score \cite{powers2020evaluation}, to exclude the presence of biases towards one type of events (see in particular Sec.~\ref{sec:stab_of_results}). Furthermore, in order to validate the generalisability of results, every training-testing process will be run using a five-fold cross validation procedure \cite{allgaier2024cross}, independently repeated three times to obtain global averages and standard deviations.

Before evaluating the performance of these models, it has to be noted that each of them is defined by a set of hyperparameters, which affect their behavior and that ought to be tuned to ensure optimal results. This is here done by resorting to \textit{Optuna} \cite{akiba2019optuna}, i.e. an automatic hyperparameter optimisation framework that combines multiple algorithms, from Gaussian process-based algorithms, to Tree-structured Parzen Estimator (TPE) and random search. For each considered ML model, the most relevant hyperparameters have been tuned using this framework, while the remainder ones are left at their default value. Each \textit{Optuna} run consists of a total of $1,200$ iterations.

Tab.~\ref{tab:optuna_accuracies_all} summarises the outcome of this process by presenting the average accuracies of the best-performing hyperparamers, for each combination of model and spatial scale. Clear differences in performance can be observed. Results are generally better for smaller spatial divisions, namely elementary blocks and sectors. Additionally, Gradient Boosting, Random Forest and Decision Tree models outperform those based on SVM, i.e. SVC and linear SVC. The highest accuracy is obtained for elementary blocks and Random Forests. Given these results, RF will be used for all subsequent analyses.

\begin{table}[!tb]
    \footnotesize
    \centering
    \begin{tabular}{|l|c|c|c|c||c|}
    \hline
    & Elementary  & Sectors & Area Control & Swedish & Model  \\
    & blocks &  & Centres & FIR & average\\\hline
        Gradient Boosting & $68.08 \%$ & $67.12 \%$  & $58.58 \%$ & $58.27 \%$ & $63.01 \%$\\\hline
         Random Forest & $70.11 \%$ & $67.80\%$ & $58.97 \%$ & $58.73 \%$ & $63.90 \%$\\\hline
         Decision Tree & $67.64 \%$ & $65.10 \%$ & $59.59 \%$ & $57.57 \%$& $62.47\%$ \\\hline
         SVC & $60.64 \%$ & $54.43 \%$ & $52.81 \%$ & $57.72 \%$& $56.40 \%$ \\\hline
         Linear SVC &  $50.29\%$ & $50.17 \%$ & $50.08 \%$ & $55.77\%$& $51.58 \%$\\\hline \hline
    Spatial region & $63.35 \%$& $60.92 \%$ & $56.01 \%$ & $57.61 \%$ & \\
    average & & & & & \\\hline
    \end{tabular}
    \caption{Average accuracies for the best-performing hyperparameters, as identified by \textit{Optuna}, for each tested ML model and each spatial scale. The last row and column respectively report the average values for respectively each spatial scale and model.}
    \label{tab:optuna_accuracies_all}
\end{table}

As a last topic, it has to be noted that the tuning has been performed independently for each spatial scale; this may introduce biases, as the best hyperparameters for, e.g., elementary blocks may not coincide with those optimal for the FIR. We here evaluate the impact of this issue by assessing the accuracy obtained by the RF model, tuned to elementary blocks, when applied to the other spatial scales. 
Tab.~\ref{tab:rf_with_volume_optimal_params} shows the outcomes of this process. Comparing these results with those of Tab.~\ref{tab:optuna_accuracies_all}, it can be appreciated that classification performances suffer almost no drop when moving away from the optimal parameter configuration \textit{Optuna} found for each spatial region. The maximum drop takes place for the FIR division, but it is still below $4\%$. Therefore, for the sake of simplicity, in what follows we will only use RF models optimised to elementary blocks. The corresponding hyperparameters are: \textit{n\_estimators}: $609$, \textit{min\_ samples\_leaf}: $3$, \textit{max\_leaf\_nodes}: $317$, \textit{max\_depth}: $309$. The best-performing hyperparameters of the rest of the models, for each of the analysed spatial divisions, can be found in App.~\ref{app:all_hyperparameters}.

\begin{table}[!tb]
    \centering
    \begin{tabular}{|l|c|c|c|c|}
    \hline
    & Elementary  & Sectors & ACCs & FIR  \\
    & blocks &  &  &  \\\hline
    Avg. accuracy & $70.11 \%$ & $67.28 \%$ & $57.80 \%$ & $54.91 \%$ \\\hline
    Acc. loss  w.r.t. & - & $0.52 \% $ & $1.17 \%$ & $3.82 \%$ \\
    RF optimal model & &  &&\\\hline
    \end{tabular}
    \caption{Classifications with sub-optimal models. Top elements report the accuracies obtained obtained for the four spatial scales and using the RF model optimised for elementary blocks; bottom ones, the corresponding drop in accuracy with respect to the use of the optimal RF model of each case - see Tab.~\ref{tab:optuna_accuracies_all}.}
    \label{tab:rf_with_volume_optimal_params}
\end{table}

\subsection{Statistical analysis of metrics' distributions}
\label{sec:stat_anal_met_dist}

One of the fundamental steps in the analyses below will be the pairwise comparison of different metrics' distributions, in order to understand how a change in the complexity of the traffic influences the probability of requiring a deconflicting action. While this comparison can be done qualitatively (i.e. graphically), we also need a quantitative method for obtaining more interpretable and rigorous results. 

Pairs of distributions will here be compared using the Wasserstein distance \cite{panaretos2019statistical}, also known as the ``Earth mover's distance''.
Intuitively, this distance measures the cost of converting, in an optimal manner, one distribution into the other, where the cost is given by the amount of probability mass transported multiplied by the corresponding distance. Note that the concepts of \textit{amount}, \textit{distance} and \textit{mass} have to be interpreted in an abstract sense, as the metric is defined on the distribution's space, where they lack the usual physical meaning. From a mathematical point of view, this distance is given by the integral:

\begin{equation}
\label{eq:wasserstein_distance}
    W_1(P, Q) = \int_{-\infty}^{\infty} \left| F_P(x) - F_Q(x) \right| \, dx,
\end{equation}

where $P$ and $Q$ stand for the distributions to be compared, and $F_i(x)$ is the cumulative distribution function (CDF) of $i$.
The implementation here used corresponds to the one available in the SciPy Python package \cite{virtanen2020scipy}. To make distances comparable across metrics with different scales, we normalise them by the pooled standard deviation, i.e. the weighted average of the standard deviations of the two distributions. This is analogous to other standardised effect sizes, such as the Cohen's $D$ \cite{cohen1988statistical}. We interpret the resulting values using conventional effect size benchmarks \cite{cohen1988statistical} (i.e. small $\approx$ 0.2, medium $\approx$ 0.5, large $\approx$ 0.8); in short, the larger the value is, the more different the two distributions are.

\section{Results}
\label{sec:results}

\subsection{Spatial and temporal effect}
\label{sec:spat_and_temp_effect}

\begin{figure}[!tb]
    \centering
    \includegraphics[width=0.8\linewidth]{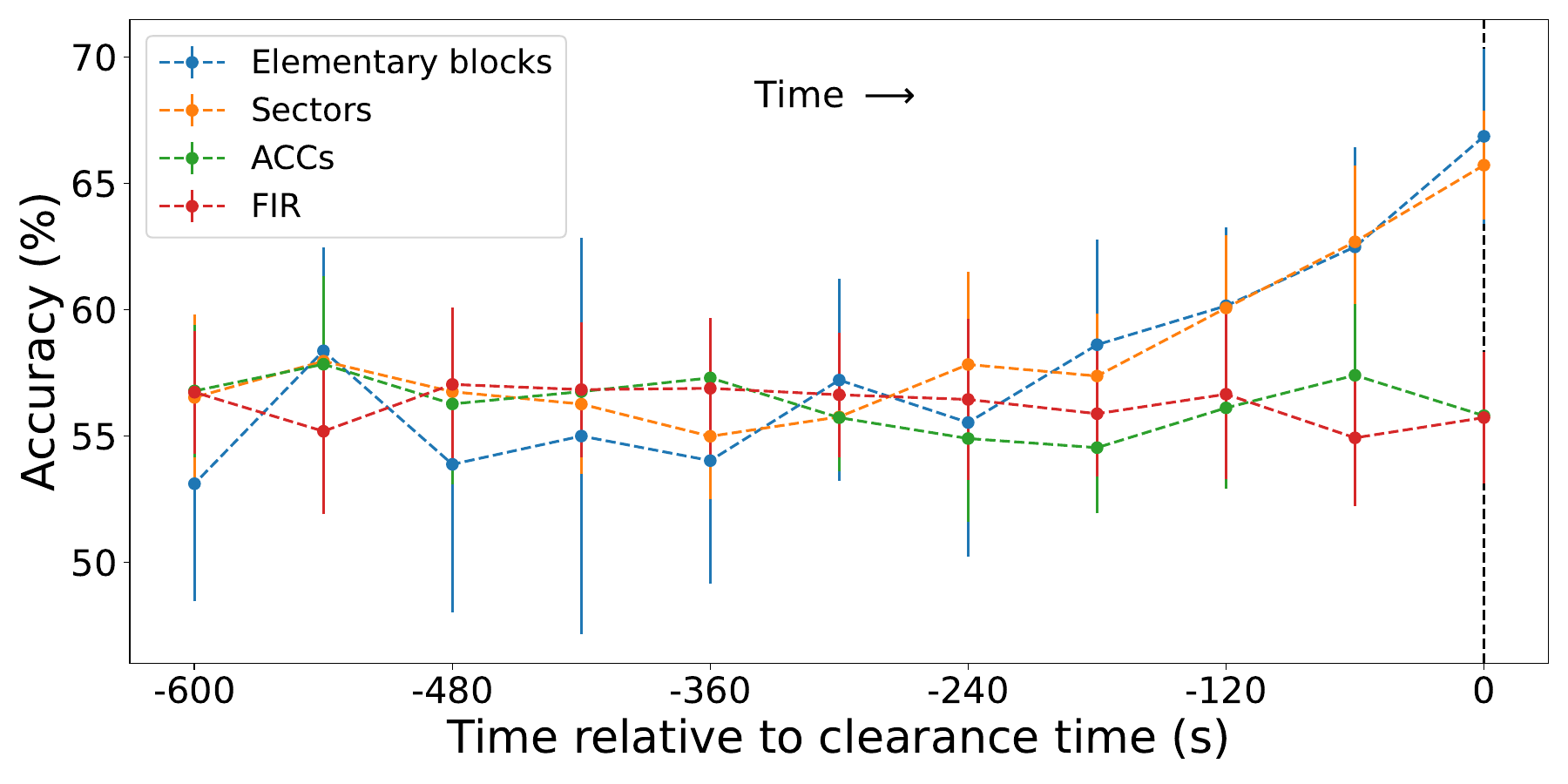}
    \caption{Evolution of the accuracy in the classification task, as a function of the spatial (different lines, see legend) and temporal (X axis) scale. Dots and solid vertical bars respectively denote the average and standard deviation. The vertical dashed black line represents the clearance time.}
    \label{fig:gradient_boost_all}
\end{figure}

We start the analysis of the results by considering how the classification accuracy evolves across spatial and temporal scales. This is depicted in Fig.~\ref{fig:gradient_boost_all}, which represents the accuracy as a function of the time offset (i.e. the time to the clearance, X axis) and of the airspace division (different lines). Results can clearly be divided in two groups: while ACCs and FIR show stable accuracies of around $55 \%$ for all values of the time offset, elementary blocks and sectors display a monotonous increase as the evaluation time gets closer to the clearance time, reaching a maximum of $\approx 67 \%$. 
It can be concluded that it is almost impossible to label events in ACCs and FIR, above what expected by chance. This may be due to the fact that not enough information about them is encoded in the features; or, conversely, that the features also include traffic that has no relation to the event under study, hence diluting the relevant information.
Consequently, in what follows, we will focus on the more informative spatial scales of elementary blocks and sectors.

\begin{figure}[!tb]
  \centering
  \includegraphics[width=\textwidth]{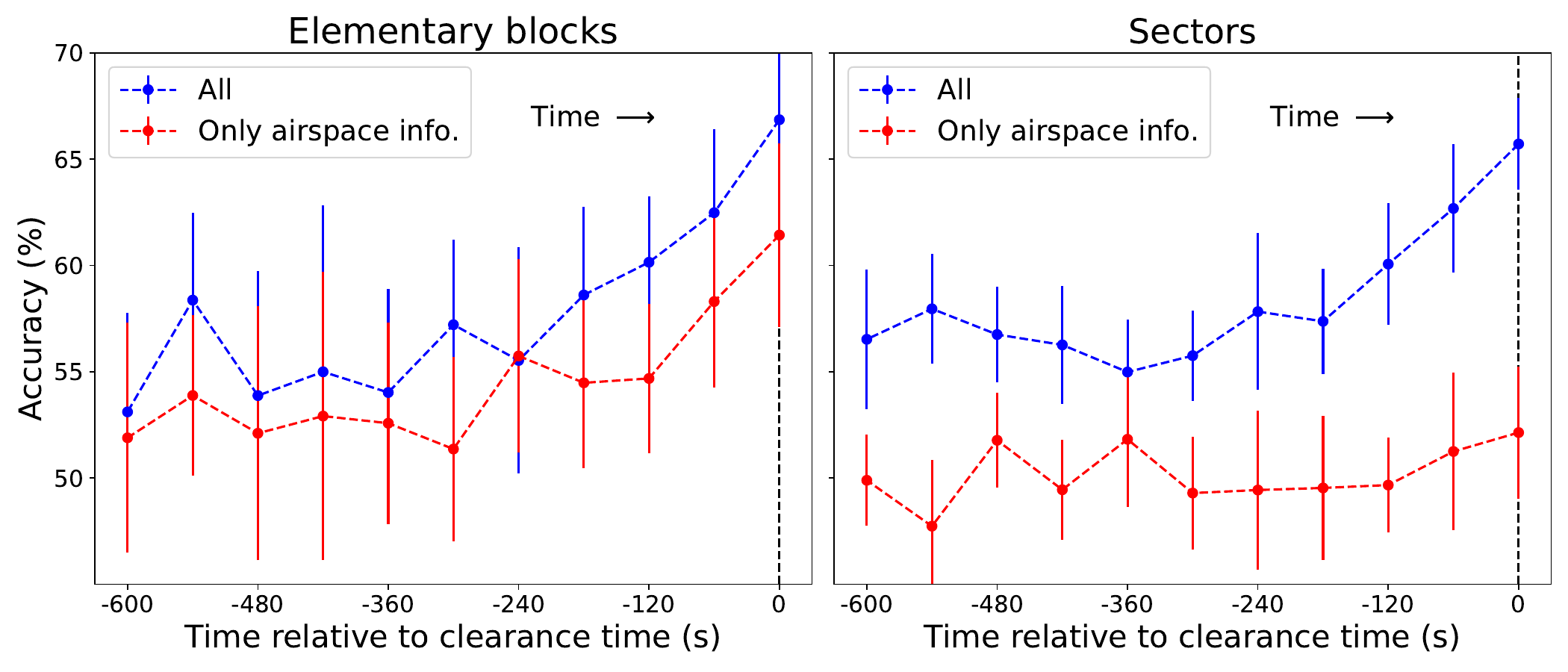}  
  \caption{ Role of airspace-related features in the classification. Each panel reports the evolution of the accuracy, as a function of the time to clearance, when models are trained with all features (blue lines) or with airspace-related features only (red lines). The left and right panels respectively correspond to results for elementary blocks and sectors, see titles. Points and solid vertical bars respectively indicate means and standard deviations. Vertical dashed black lines represent the clearance time.}
  \label{fig:area_score_evolution}
\end{figure}

We further check whether the obtained classification scores can be explained by using information about the airspace structure alone. Towards this end, we executed the same classification task as in Fig. \ref{fig:gradient_boost_all}, but only including \textit{Sector Area} and \textit{Sector ID} as inputs. Results, reported in Fig. \ref{fig:area_score_evolution}, highlight an important difference: while the performance of the classification task is qualitatively the same in the case of elementary blocks (see left panel), it drops to $\sim 50\%$ accuracy for sectors (right panel).
This suggests that the classification model is mainly learning from the characteristics of each elementary block, while the rest of the complexity metrics appear to be redundant. 
The reason behind this behaviour can be traced back to the high heterogeneity of elementary blocks, see App. \ref{app:ind_sec_vol} for an in-depth analysis. Due to the lack of relevance of the complexity metrics at such smaller scale, and in line with the objective of this work, analyses below will only focus on sectors.

\subsection{Impact of representation and classification choices}
\label{sec:stab_of_results}

While the scores presented in the previous section seem to indicate that a classification above randomness can be achieved, these may also be the result of biases arising from the way the airspace is described - as partly seen in Sec. \ref{sec:spat_and_temp_effect} in the case of the spatial scale. In order to exclude this, we here check the impact of some representation and classification choices on the process.

\begin{figure}[!tb]
    \centering
    \includegraphics[width=\linewidth]{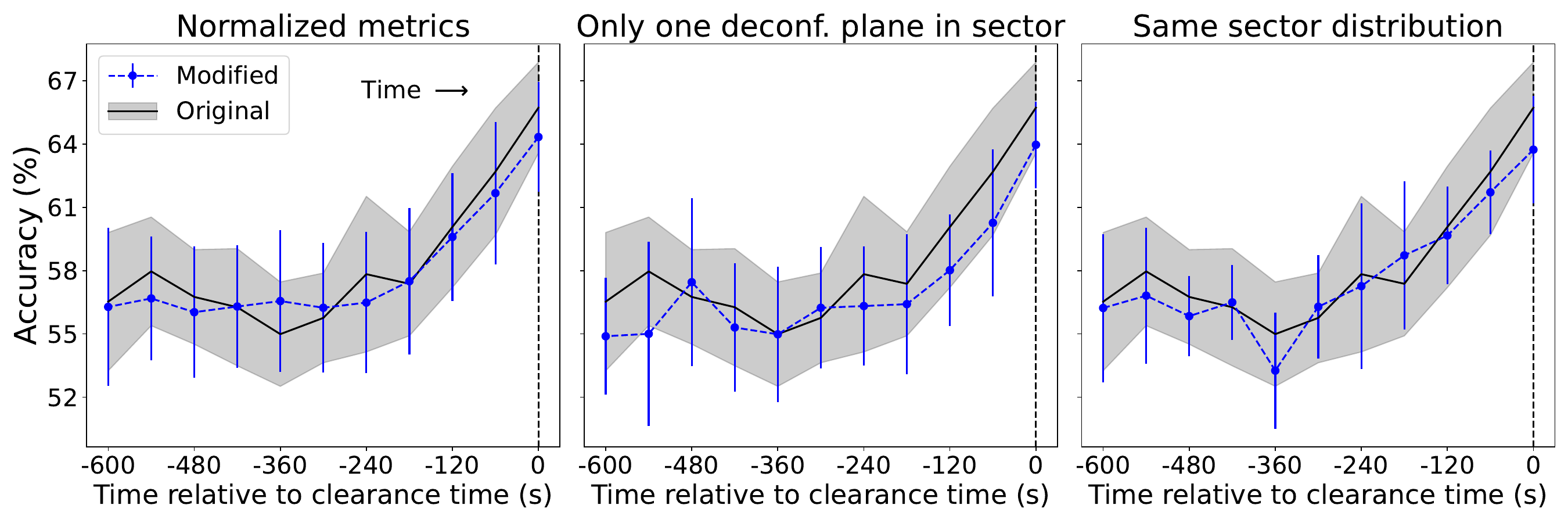}
    \caption{Evolution of the average accuracy of the model as a function of the time to clearance, using normalised metrics (left), events in which only one of the two aircraft is in the analysed sector (center), and enforcing equal distribution of both types of events across all sectors (right). Black solid lines and grey bands correspond to the original results of Fig. \ref{fig:gradient_boost_all} for sectors. Vertical dashed black lines represent the clearance time. See main text for details. }
    \label{fig:different_normalizations_score_evolution}
\end{figure}

We start by firstly considering three traffic representation alternatives, whose results are respectively depicted in the three panels of Fig. \ref{fig:different_normalizations_score_evolution}:

\begin{enumerate}
    \item Metric normalisation (left panel). The complexity metrics describing the traffic are normalised according to the traffic volume and the geometry of the sector, when relevant. This is done to assess whether, after eliminating these two aspects, the previously obtained classification is maintained; or, on the contrary, these aspects are dominating the results.

    \item Presence of the conflicting aircraft (central panel). Especially in cases in which only a handful of aircraft are present in the considered airspace, it can be argued that the conflicting pair may have a disproportional impact in the complexity metrics - i.e. that we may be describing their relation, as opposed to the full traffic. We thus test the classification using only snapshots in which one aircraft is not in the airspace under analysis - note that the other one, i.e. the aircraft to which the deconfliction is issued, must be present at the time of the event by construction.
    
    \item Distribution of conflict-free events (right panel). By default, conflict-free events are identified over the whole airspace and then randomly included in each classification - see Sec. \ref{sec:conflict_free}. It may thus happen that one specific sector is over-represented, especially if it contains few crossing aircraft, and hence less opportunities for conflicts. We here thus considered a strict sampling, in which the number of both types of events is kept equal for all sectors. 
\end{enumerate}

In all three cases, the obtained classification accuracy is similar to that obtained without any modification, thus indicating that these analysis choices have a negligible impact in the analysis. Results in the left panel suggest that the number of aircraft and the sector geometry have little impact; hence, that the detected patterns are robust against them and are generalisable. This does not exclude the influence of these properties in other complexity metrics; for a full analysis of the correlation between them, see App. \ref{app:metrics_correlation}.
The central panel of Fig. \ref{fig:different_normalizations_score_evolution} further highlights an interesting fact: the future need of a deconfliction action is detectable even before the involved aircraft have entered the airspace under analysis. In other words, such need is not due to these two aircraft and the corresponding conflict geometry; but is rather the result of how the traffic as a whole is evolving. The implications of this result will become clearer below.

\begin{figure}[!tb]
    \centering
    \includegraphics[width=0.66\linewidth]{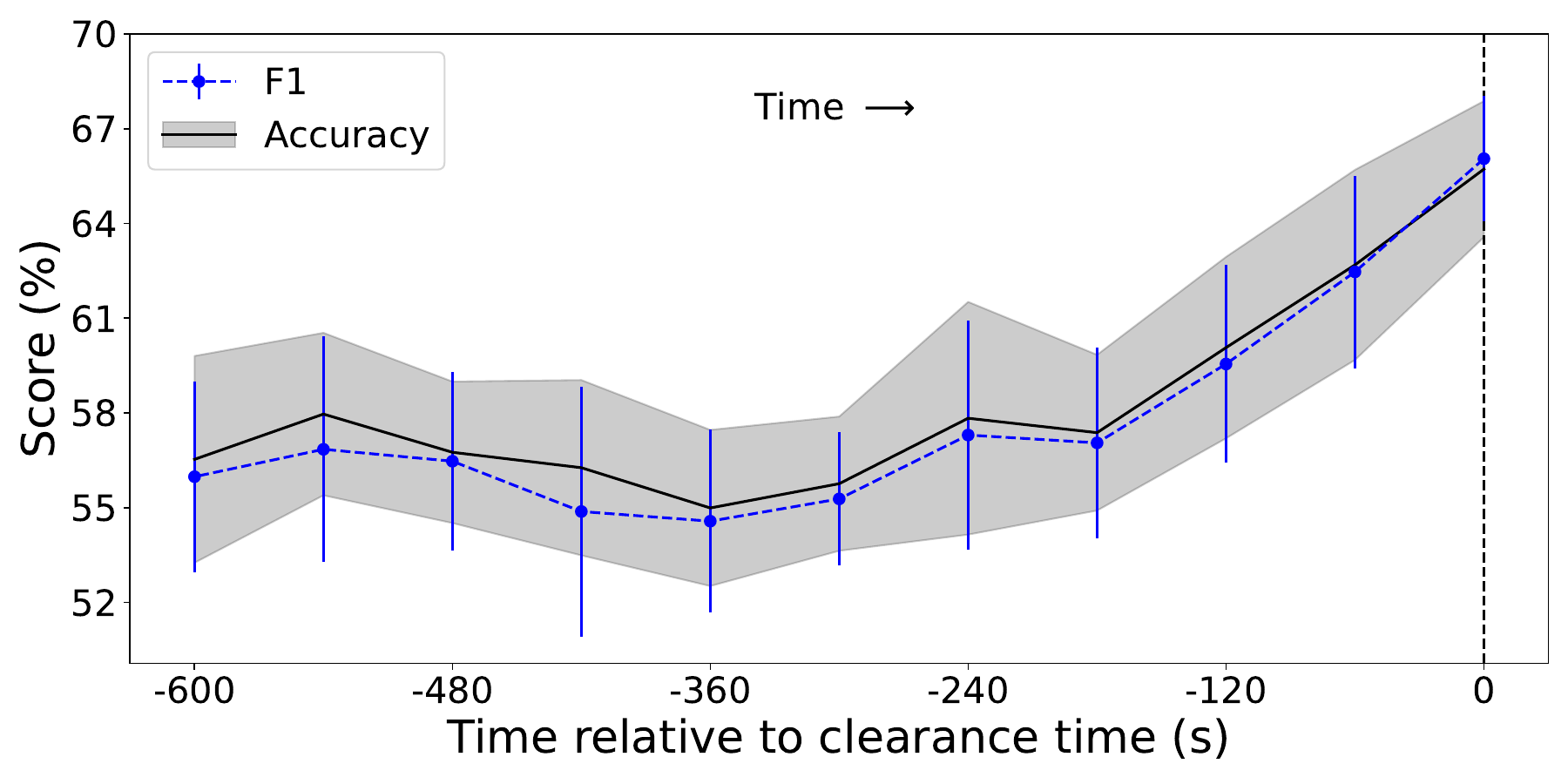}
    \caption{F1 score as a function of the time to clearance; blue dots indicate the average values, and the vertical lines the corresponding standard deviation. For the sake of comparison, the accuracy of the model is reported by the solid black line (average) and the grey band (standard deviation). Vertical dashed black line represents the clearance time.}
    \label{fig:f1_scores}
\end{figure}

Secondly, we analyse whether the performance of the models may be asymmetrical between deconfliction and conflict-free events - as the accuracy equally weights both classes. For this we resort to the F1 score, which qualitatively is a measure of the accuracy over the deconflicting events only. As can be seen in Fig.~\ref{fig:f1_scores}, both score metrics yield very similar results, and all the average values of the F1 scores are within the standard deviation range of their accuracy counterpart; in short, the classification performance is the same for deconflicting events and conflict-free ones.

\subsection{Metrics' distributions comparison}
\label{sec:metrics_distributions_comparisons}

The result presented in the right panel of Fig.~\ref{fig:area_score_evolution}, back in Sec.~\ref{sec:spat_and_temp_effect}, suggests two complementary views to what characterises a deconfliction event. Firstly, the maximum accuracy, and hence the maximum difference between types of events, is observed at clearance time. Secondly, the accuracy gradually improves when we approach that time; the system seems therefore to be evolving, from a normal to an altered state. These two views can naturally be tested using the Wasserstein distance, as previously defined in Sec.~\ref{sec:stat_anal_met_dist}. Specifically, two comparisons ought to be performed. On the one hand, one may quantify the differences between the two types of events, by comparing the values of each complexity metric, at clearance time, between deconfliction and conflict-free events. This captures which complexity metrics are relevant to detect a deconfliction event. On the other hand, we further compare the values of the metrics for deconfliction events only, at the time of clearance vs. eight minutes before. This second analysis thus describes how the traffic patterns evolve when approaching the altered state. Additionally, and for the sake of completeness, we further evaluate a third option: the distance, for conflict-free instances, between metrics at the time of the event and eight minutes before. 

From a data analysis perspective, if the ML model is able to discriminate between the two types of events at the time of clearance, some of its input features (here, some of the complexity metrics) must be different, hence their distribution should have a significantly positive $W$ (see Eq.~\ref{eq:wasserstein_distance}). Additionally, if an evolution is indeed present when approaching a deconfliction event, the same must also be true when considering the status of the system eight minutes before. Finally, differences should not be observed in the case of conflict-free instances - as the clearance time is in this case arbitrary, hence the system should not display any consistent evolution.

Tab.~\ref{tab:significant_metrics_distributions_differences} presents the results for metrics that yielded, in at least one of the analysis, a distance $W \geq 0.2$. For the sake of completeness, full results are reported in App.~\ref{app:met_dist_diff}.
As previously hypothesised, moderate values of $W$ are observed for several complexity metrics. Most notably, $W$s in the first two comparisons are qualitatively similar, as are the directions of the metrics changes, as reported in the fifth column, \textit{Tendency} (see the table caption for its definition). 
In other words, the difference between deconfliction and conflict-free events, at the time of clearance, is similar to that between the deconfliction event itself and eight minutes before it. Even in the two cases not reaching a $W$ of $0.2$, namely the \textit{number of aircraft accelerating} and the \textit{occupancy variation}, values are close to that reference, and are one order of magnitude larger than what obtained in the comparison of conflict-free events (last column). This indicates that conditions leading to a deconfliction are not static, but rather build up in the minutes prior to the event itself.
The same is not true in the case of conflict-free events, confirming that the observed evolution is something specific to situations leading to a deconfliction.

\begin{table}[!tb]
\centering
\resizebox{\linewidth}{!}{
\begin{tabular}{l l c c c | c}
\hline
        Group & Metric & Deconflictions  & Clearance time & Tendency & Clearance time\\
                & & vs conflict free & vs 8 min.  & & vs 8 min.\\

        & & at clearance & before.  & & before.\\
                & & time (W) & Deconf. (W) & & Conf. free (W)\\

\hline
Trajectories' &  Mean distance flown in sector & $0.273$ & $0.245$ & $\downarrow$ & $0.064$  \\
characteristics&  Mean time flown in sector & $0.251$ & $0.256$ & $\downarrow$ & $0.048$\\
& Number of aircraft close & $0.311$ & $0.297$ & $\uparrow$ & $0.034$ \\
& to sector border & & & & \\
\hline
Individual  &Number of aircraft with absolute & $0.387$ & $0.309$ & $\uparrow$ & $0.024$
\\
 aircraft  & ground speed change $>$ 10 kt in 2 min & & & & \\
 characteristics & Number of aircraft accelerating & $0.215$ & $0.153$ & $\uparrow$ & $0.038$ \\
\hline
Occupancy & Number of aircraft & $0.289$ & $0.270$ & $\uparrow$ & $0.064$\\
\hline
Occupancy & Occupancy variation & $0.200$ & $0.168$ & $\uparrow$ & $0.091$\\
Variation & Number of entering aircraft & $0.352$ & $0.209$ & $\uparrow$ & $0.034$\\
 & Number of exiting aircraft & $0.258$ & $0.354$ & $\uparrow$ & $0.052$\\
\hline
 & Number of pairs with reduced & $0.254$ & $0.294$ & $\uparrow$ & $0.011$\\
 & separation & & & & \\
 Pairwise & Number of potential crossings & $0.339$ & $0.336$ & $\uparrow$ & $0.021$ \\
relations & Number of potential crossings & $0.283$ & $0.305$ & $\uparrow$ & $0.023$\\
& with convergence angle $>20^{\circ}$ & & & & \\
& Number of potential conflicts &  $0.353$ & $0.318$ & $\uparrow$ & $0.063$ \\
& Number of potential conflicts & $0.283$ & $0.254$ & $\uparrow$ & $0.037$\\
& with convergence angle $>20^{\circ}$ & & & & \\
& Pairwise horizontal& $0.266$ & $0.244$ & $\uparrow$ & $0.060$ \\
&  distances entropy & & & & \\
\hline
\end{tabular}
}
\caption{Wasserstein distance values $W$ for the three comparisons described in the main text. \textit{Tendency} column reports the sign of the change in the mean metric value for deconfliction snapshots at clearance time, relative to both conflict-free snapshots at clearance time and deconfliction snapshots 8 min before clearance time (i.e. $\uparrow$ positive change; $\downarrow$ negative change). Only metrics with $W \geq 0.2$ for at least one of the comparison are included. The data set used is that considering all the scenarios with non-normalised metrics.
}
\label{tab:significant_metrics_distributions_differences}
\end{table}

\begin{figure}[!tb]
    \centering
    \includegraphics[width=\linewidth]{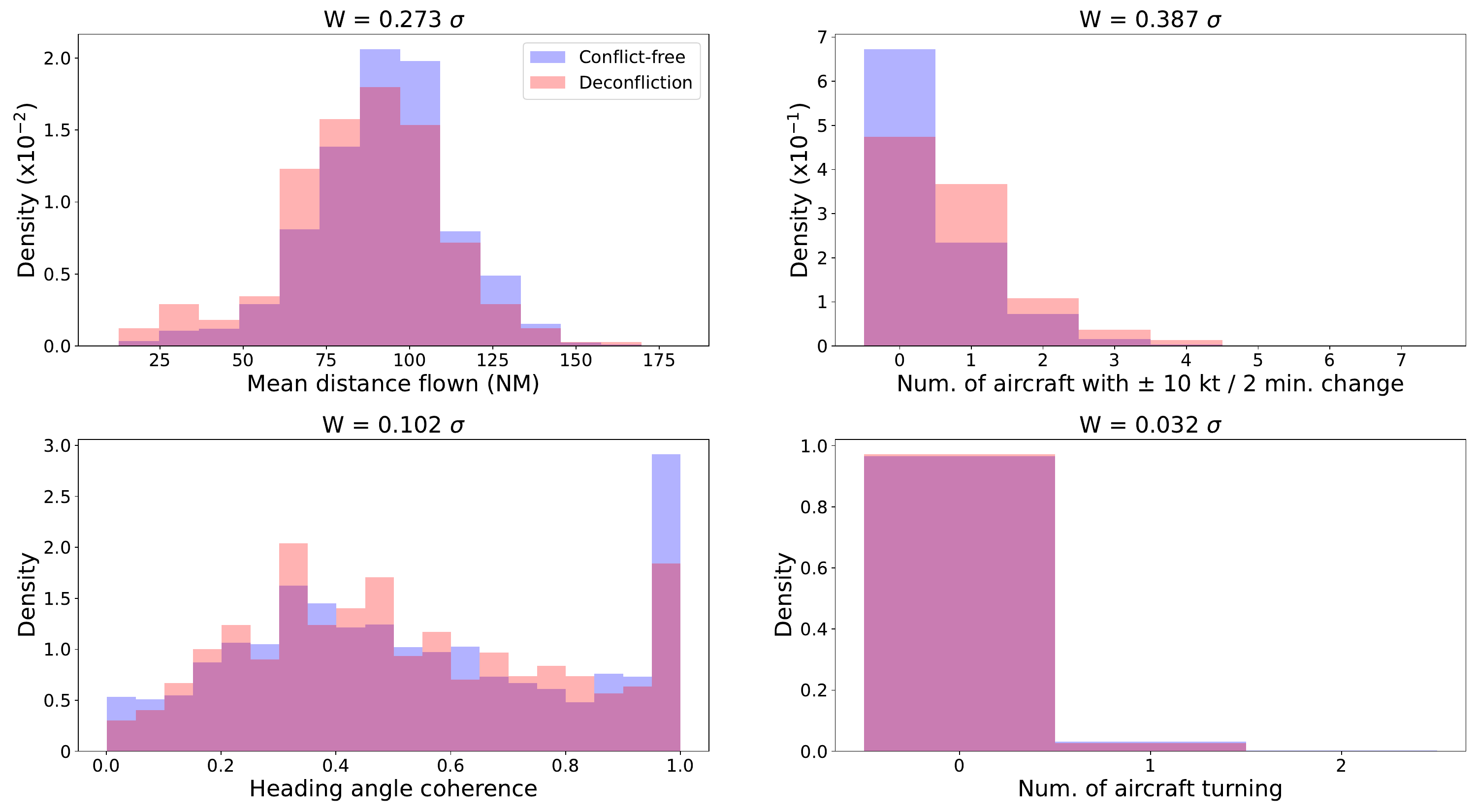}
    \caption{Histograms comparing the metrics' distributions between deconflicting and conflict-free events when evaluated at clearance time, for a total of four different metrics: mean distance flown in sector (top left), number of aircraft which change their speed in at least $10$ kt in the following $2$ min. after evaluation time (top right), heading angle coherence (bottom left) and number of aircraft turning at the moment of evaluation (bottom right). The corresponding Wasserstein distance is reported on top of each panel.}
    \label{fig:severe_vs_no_severe_clearance_time_distributions}
\end{figure}

For the sake of illustration, Fig.~\ref{fig:severe_vs_no_severe_clearance_time_distributions} reports the histograms for some metrics comparing deconfliction and conflict-free scenarios at clearance time. The top panels show examples of two metrics for which the Wasserstein distance identified an effect larger than small, i.e. greater than $0.2$. 
Conversely, the bottom right panel reports the distribution for a metric (i.e. the {\it number of aircraft turning}) for which no differences can be detected.
Finally, the bottom left panel shows the pathological case of the heading angle coherence: despite $W$ being small, differences between events can still be identified. Specifically, the peak on $1$ observed for conflict-free snapshots disappears for deconfliction ones - i.e. deconflicting events are usually not associated with traffic patterns in which all, or almost all, aircraft have the same heading (see App.~\ref{app:compl_metrics} for the corresponding definition of this metric). While this difference is not large enough to be captured by the Wasserstein distance, ML models can still leverage this information.

\subsection{Time evolution analysis}
\label{sec:time_evol_anal}

The results presented up to here highlight an interesting point: the need of a deconfliction action does not appear suddenly, but can rather be predicted a few minutes before the action is taken. In other words, the traffic structure evolves through time towards an altered state in which the intervention is required. 
We here study more in depth this aspect, i.e. the evolution of the model's accuracy as the evaluation gets closer to the clearance time.

\begin{figure}[!tb]
    \centering
    \includegraphics[width=0.66\linewidth]{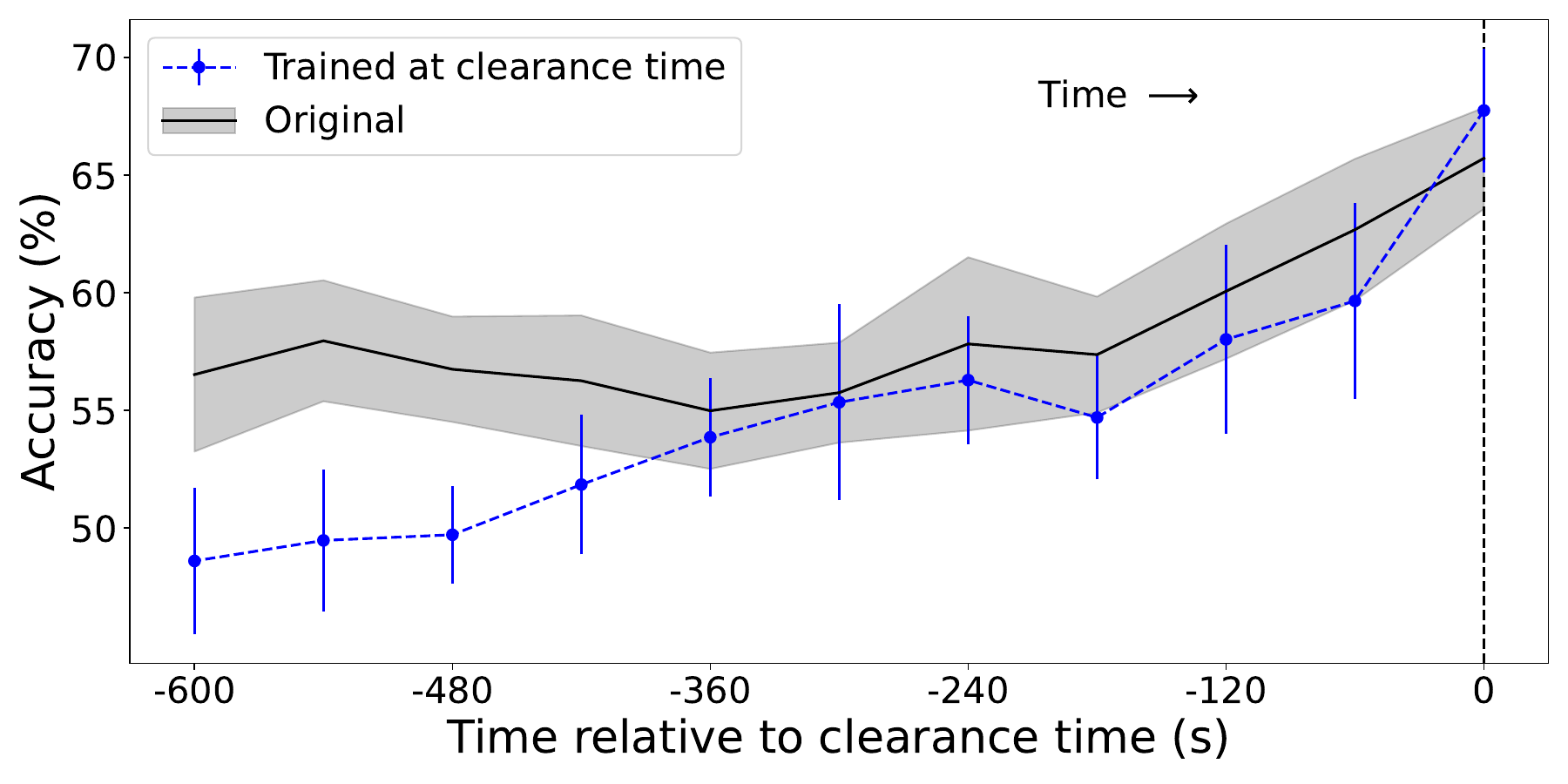}
    \caption{Average accuracy as a function of the time, relative to clearance, to which the data used to test belong to, when the model used for all the evaluations is that trained with data associated to clearance time (blue line). Black line and the shadow area around it correspond to the average accuracy and standard deviation, respectively, of the original model.  The vertical dashed black line represents the clearance time.}
    \label{fig:clearace_time_trained_model_scores}
\end{figure}

We firstly analyse the average accuracy of a model trained at clearance time, when tested with data from different timestamps before that moment. This allows to assess whether the patterns observed at the time of the deconfliction are also present before it.
Fig.~\ref{fig:clearace_time_trained_model_scores} presents the results. 
The accuracy is qualitatively the same, irrespectively on which data the model is trained; only beyond $420$ seconds the modified model presents a worse accuracy. 
In other words, in the six minutes prior to the clearance time, the difference between events is encoded in the same metrics and in similar patterns, while for earlier times this is no longer the case. Beyond these six minutes, an improvement of the accuracy of $\sim 5\%$ (see the distance between both lines in the left part of Fig.~\ref{fig:clearace_time_trained_model_scores}) can thus only be achieved by accounting for different patterns in the data that are specific to that temporal horizon.

We next focus on the deconflicting events that are correctly classified at clearance time; more specifically, on those events that the model (trained at the same moment of the deconfliction) is able to correctly label in all five cross-validation folds. 
This guarantees that the classification is robust; or, alternatively, that the model is highly confident in the assigned label. 
We next track how those events were classified by the models trained before the deconfliction events, and calculate the earliest time at which they were correctly categorised; and the last time stamp with a correct classification, i.e. the earliest they have been correctly predicted.
In short, we try to model how long in advance the models were able to understand their nature, i.e. the lead time; and, from an operational point of view, how long in advance can deconfliction events be predicted, at least according to the ML model. 

\begin{figure}[tb!]
    \centering
    \includegraphics[width=\linewidth]{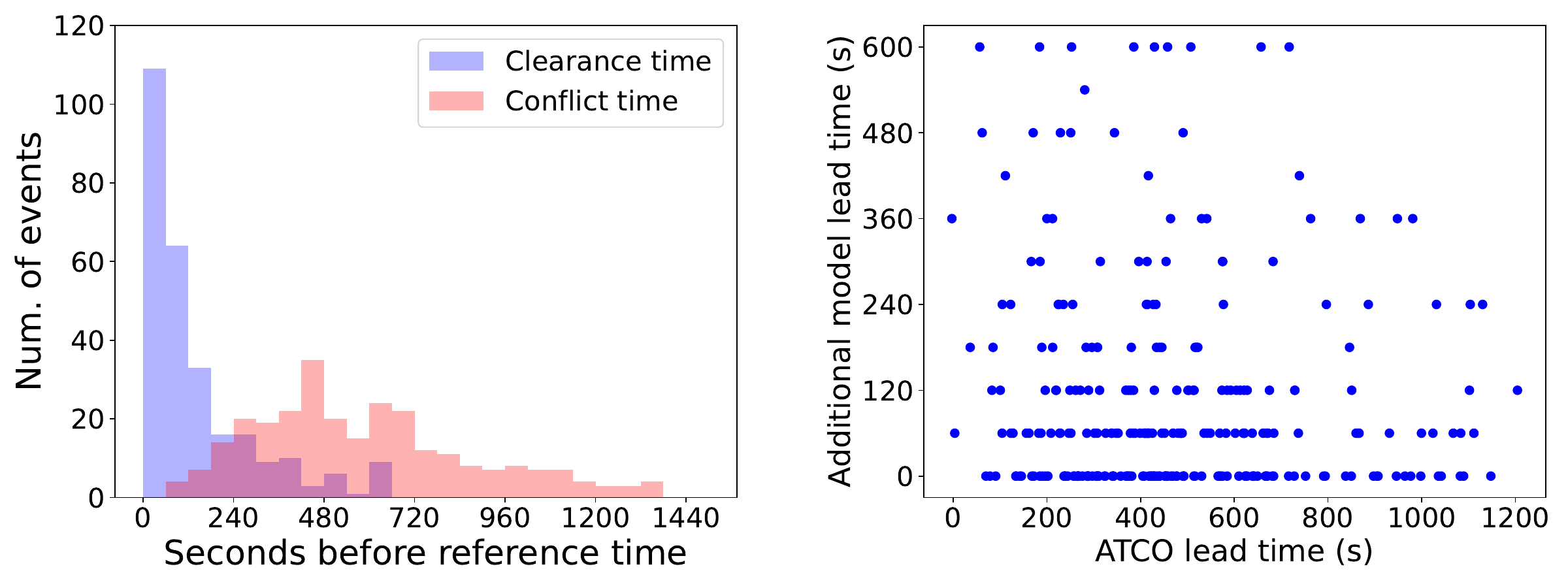}
    \caption{Lead time to classification. Left panel shows the number of deconfliction events correctly classified as a function of the earliest correct classification time, calculated with respect to both the respective clearance (blue) and conflict (red) time. Right panel reports a scatter plot of the additional model lead time as a function of the ATCO lead time, both in seconds, for each individual considered deconfliction event. See main text for definitions. }
    \label{fig:last_time_to_predict}
\end{figure}

The left panel of Fig.~\ref{fig:last_time_to_predict} shows the results of this analysis, in terms of the number of deconflictions correctly predicted as such as a function of the time before the clearance (see the blue bars). Most of the events are labelled as deconflictions only within two minutes of the clearance. The red bars reports the same information, but considering as reference the time of the predicted conflict itself (see Sec.~\ref{sec:operational_data}); the peak is moved to the central part of the distribution, more specifically to a value of around $8$ minutes. For the sake of completeness, the right panel shows a scatter plot of the additional lead time of the model against the ATCO (i.e. how much in advance it can classify the scenario), as a function of the ATCO lead time (i.e. the conflict time minus the clearance time).

Interpreting these results from the viewpoint of the management of the traffic, the future appearance of a conflict, and hence the need of a deconfliction, can usually be detected well in advance. For a ML model, this lead time is on average of $8$ minutes; and notably, this data-based identification happens up to two minutes before the ATCO intervenes in the system. Thus, while ML models can identify patterns in airspace complexity that may escape the human assessment, the human performance is not far behind. Also, when the human decision happens close to the conflict, the ML model is not able to significantly improve the lead time; as illustrated in the scatter plot, while some points with a high additional lead times are present, the bulk is concentrated in the bottom part of the graph; late decisions are thus generally structural, and not due to human inefficacy.

\section{Discussion}
\label{sec:discusion}

Reviewing the results, the first one that has to be discussed is the fact that a classification is, in fact, possible. Accuracies over $65\%$ indicate that, in general, deconfliction events occur under specific airspace situations and traffic patterns, which can be identified and highlighted. 
Note, nonetheless, that a specific spatial division must be used to this end. If large areas of airspace are evaluated (see FIR and ACCs in Fig.~\ref{fig:gradient_boost_all}), irrelevant traffic hides and averages out the characteristics proper of the deconfliction event. On the other hand, if too reduced areas are considered (see left panel of Fig.~\ref{fig:area_score_evolution}), relevant information is lost. Not surprisingly, sectors, being the natural spatial working scale of ATCOs, are the optimal scale.

The comparison here proposed naturally supports the identification of the specific characteristics of the airspace that differ between the two types of scenarios, and hence contribute to the identification of a future deconfliction event. Tendencies reported in the fifth column of Tab.~\ref{tab:significant_metrics_distributions_differences} (see the table caption for their definition) show that, for deconflicting events close to clearance time, the number of aircraft that are inside the sector is generally higher. This is in agreement with the well-established traffic count as the main indicator for restricting airspace capacity, as currently used in real operations; at the same time, this result acts as a sanity check for this methodology, and only part of the classification accuracy comes such metric (see the  discussion in Sec.~\ref{sec:stab_of_results}). Less trivial results show that deconflicting events tend to take place in situations in which aircraft are flying close to the borders of the sector, or equivalently, when aircraft fly inside it for short times and distances. This suggests that entering/exiting traffic is more challenging to manage by the ATCOs, hence leading to more complex situations. This is further supported by higher entrance/exit rates and occupancy variations. Additionally, results suggest that aircraft not established en-route are more difficult to control, as higher number of accelerating aircraft and higher absolute values of speed changes are observed for deconfliction events. Finally, a significant impact of pairs with reduced (and reducing) separation is observed; as the number of such pairs increases, so does the probability that the ATCOs is not able to optimally deal with all of them. While these aspects have previously been reported in the literature, as they align with the cognitive idea of a higher workloads, we here prove their importance in the appearance of deconfliction events in a numerical and objective way.

While these results may seem prima facie minor, the proposed approach highlights two aspects that have seldom been described in the literature.
Firstly, the information to predict the need of a future deconfliction action is encoded in the traffic, and not in the conflicting pair. As shown in Sec. \ref{sec:stab_of_results}, the two types of instances can be classified even before both involved aircraft have entered the airspace under analysis. This is a radical shift from the usual conception of safety events: they seem to be caused not only by an abnormal dynamics of the two involved aircraft, but also by the structure created by all flights together. The latter may be requiring the attention of the ATCO, who then may fail to identify a problematic situation that is evolving. It may even be hypothesised that, had the second conflicting aircraft not entered the airspace, a different conflict might have developed. 

Secondly, the need of a deconfliction action is not something instantaneous, but rather the end of a dynamical process that spans several minutes (on average, eight minutes before the conflict itself, see Fig. \ref{fig:last_time_to_predict}). This confirms the idea of something changing in the traffic as a whole, i.e. a progressively degrading situation, that the ATCO cannot recognise straight away; and is further aligned with Dekker's ``Drift into Failure'' framework, which argues that safety events are not discrete instants but the culmination of accumulated latent conditions \cite{dekker2011drift}. 

Taken together, these two novel conclusions support the possibility of developing early warning systems. We have statistically proven that degraded situations are independent of the specific aircraft generating the conflict, i.e. a general warning may be raised about the status of the airspace as a whole; and that they further are antecedents of the events themselves, i.e. a warning can be raised before the safety event materialises. By being able to integrate more information, ML models here considered are slightly better than the human ATCOs, being able to detect some degraded situations with two minutes of additional lead time (see Fig. \ref{fig:last_time_to_predict}).

In spite of these interesting results, several limitations of this work have to be highlighted. Firstly, the dataset here analysed does not focus on unexpected low-frequency events, e.g. weather avoidance or pilot non-compliance \cite{nilsson2023swedish}. The ``path to degradation'' previously described would not apply in the case of abrupt events, as e.g. of an aircraft suddenly starting descending because of a power loss. Additionally, it also presents a limited spatial and temporal coverage; and further lacks other relevant information, as e.g. the list of open sectors. In short, the characteristics of the data set lead to the second limitation: the limited generalisability of the results. As a third point, extreme care has been exercised in the analysis of the impact of several methodological choices, as e.g. how metrics are defined and how models are tuned. Still, some of them could be improved in the future. To illustrate, the hard threshold between deconfliction vs. conflict-free could be made more flexible; more airspace complexity metrics could be added; and even some metrics could be redefined (as e.g. the entropy is a biased statistical indicator). Finally, the attentive reader would have noted that a classification accuracy of almost $70\%$ is in general not enough to support an early warning system, as too many false positives would be triggered.

These limitations well support future expansions of the work here proposed. On the one hand, classification accuracies could be improved by using more sophisticated models, such as Deep Learning ones; while these usually display higher performance, they also come at the cost of a reduced explainability. On the other hand, given that the evolution of the traffic patterns has been found to be a key element (see Sec.~\ref{sec:metrics_distributions_comparisons}), the evolution of the complexity metrics, as opposed to scalar values, could be considered. Combining both, the optimal solution may reside in the use of Deep Learning models with attention mechanisms.
The generalisability of the results could be validated by repeating the same analysis in other regions of the airspace; while not common in the open domain, data about deconfliction events in other regions of Europe are available \cite{gaume2025extracting}. Finally, to add interpretability to the results, a key point would be their evaluation by ATCOs and air traffic experts, to see if they are aligned with what is daily observed in a real operational environment.

\section{Conclusions}
\label{sec:conclusions}

This work presents, to our knowledge, the first deconfliction vs. conflict-free comparison of air traffic trajectories based on data analyses. It thus goes beyond the state of the art on air traffic management, by allowing to rate how meaningful a property (here, a complexity metric) is, rather than just describing a deconfliction \cite{gaume2025extracting}.
Specifically, we have characterised the evolution of the traffic structure around events that lead to deconflicting actions, and compared it to that of events free from conflicts. We described such structures through complexity metrics measured at different times before the clearance instruction and in different spatial regions, and used these as features for machine learning models. The resulting classification score was finally used to concluded that deconfliction events evolve differently than conflict-free situations, especially when approaching the clearance time. Additionally, the optimal temporal and spatial scales for the observation have been identified (Sec.~\ref{sec:spat_and_temp_effect}); results have been shown to be robust to external factors, such as traffic loads or differences in sectors (Sec.~\ref{sec:stab_of_results}); the specific characteristics of airspace that are the main drivers of the differences in evolution have been identified (Sec.~\ref{sec:metrics_distributions_comparisons}); and an in-depth analysis of the temporal evolution of deconfliction events has been performed (Sec.~\ref{sec:time_evol_anal}).

In synthesis, this work presented a first numerical and data-based methodology to study deconfliction events; by comparing them with conflict-free scenarios, to understand what conditions are behind their appearance; and to describe the time-dependent degradation of the system. The identification of such degraded states, before the safety hazard is actually detected by the ATCOs, is of high relevance to the air traffic management research community, as these insights can be useful in the design of safer and more efficient operations.

\section*{Acknowledgements}

R.L.-M. and M.Z. acknowledge support from the María de Maeztu project CEX2021-001164-M funded by the MICIU/AEI/10.13039/501100011033. R. L.-M. acknowledges support from the Spanish Ministry of Science, Innovation and Universities through the grant FPU22/03765.

\clearpage
\newpage

\appendix

\section{Complexity metrics}
\label{app:compl_metrics}

We here report a detailed explanation of the sector complexity metrics used throughout this work, as previously introduced in Sec.~\ref{sec:complexity_metrics}. The interested reader will find additional information about each metric's concept in the corresponding references. Additionally, App. \ref{app:norm_metrics} and App. \ref{app:metrics_correlation} will respectively discuss their normalisation, as used in Sec.~\ref{sec:stab_of_results}, and their pairwise correlations.

\begin{itemize}
    \item \textit{Mean horizontal distance} \cite{chatterji2008analysis}: Average horizontal distance across all pairs of aircraft, in NM.
    
    \item \textit{Mean vertical distance} \cite{chatterji2008analysis}: Average vertical distance across all pairs of aircraft, in FL (hundreds of ft).

    \item \textit{Pairwise horizontal distances entropy} \cite{stein1985workload, histon2002structural}: Shannon's entropy of the probability distribution of the horizontal distances between pairs of aircraft, measured in NM and grouped into bins of width 5 NM. The value is thus proportional to the heterogeneity observed in the horizontal separations between aircraft, and can further be understood as a proxy of the concept of aircraft clustering on an airspace's region.

    \item \textit{Number of aircraft close to sector border} \cite{prandini2011toward}: Number of aircraft located within $10$ NM of the nearest point of the boundary of the airspace region under analysis.

    \item \textit{Mean traffic convergence/divergence} \cite{delahaye2000air, xiao2016atcem}: Measure quantifying the tendency of traffic to become more concentrated or dispersed over time. It is calculated by subtracting the mean pairwise horizontal distance at the event timestamp from the value calculated at the next available timestamp. Negative values denote convergence, while positive values denote divergence. Larger absolute values indicate stronger convergence or divergence trends.

    \item  \textit{Number of pairs with reduced separation}\cite{eurocontrol2000complexity}: Number of pairs of aircraft that are under reduced separation at evaluation time. For this work, two aircraft are considered to be under reduced separation if they fly within $10$ NM horizontally and $2,000$ ft vertically, simultaneously \cite{lopez2023propagation}.

    \item  \textit{Number of different flight levels in use} \cite{moreno2022determination}: Number of different flight levels in which at least one aircraft is flying, within the airspace region under analysis. With the goal of capturing the airspace's flight level structure, each aircraft's altitude is rounded to the closest multiple of $10$ FL.

    \item  \textit{Flight levels occupancy entropy} \cite{majumdar2002factors}: Measure proportional to the heterogeneity in the distribution of aircraft across flight levels. It can be interpreted as a proxy for the vertical concentration of traffic. With the goal of capturing the airspace's flight level structure, each aircraft's altitude is rounded to the closest multiple of $10$ FL; the Shannon's entropy of the resulting distribution is then calculated. Unlike the \textit{Number of different flight levels in use}, this metric additionally weights each occupied level by the number of aircraft in it.

    \item  \textit{Heading angle coherence} \cite{brinton1996dynamic}: Measure of the mix of aircraft's heading angles within the spatial region under analysis, such that smaller values indicate a greater diversity of flight directions. To quantify this metric, the phase-coherence value as defined by the respective order parameter ($r$) in the Kuramoto model is used \cite{acebron2005kuramoto}, assuming a population of oscillators whose phases are the heading angles. Mathematically, its definition is:

    \begin{equation}
        r(t) e^{i\psi(t)} = \frac{1}{N} \sum_{j=1}^{N} e^{i\theta_j(t)},
    \end{equation}

    where $t$ is the evaluation time; $\theta_j(t)$ is the heading of aircraft $j$ measured in radians; $r(t)$ is the coherence level; and $\psi(t)$ is the average heading across all aircraft. The more similar headings the aircraft have, the larger the value will be, with $1$ meaning equal heading for all aircraft; conversely, smaller values indicate a large heading mix, with $0$ indicating a uniform distribution of headings. This specific way of measuring this characteristic is here used as it correctly eliminates the angles' range discontinuities.

    \item  \textit{Mean ground speed} \cite{xiao2016atcem}: Average ground speed across all aircraft, in knots.

    \item  \textit{Ground speeds variance} \cite{xiao2016atcem}: Standard deviation of the aircraft's ground speeds distribution.
    
    \item  \textit{Mean vertical speed} \cite{gianazza2006selection}: Average absolute vertical speed, in FL per seconds, across all aircraft within the airspace region under analysis. Taking absolute values ensures that the metric reflects the magnitude of the altitude changes, rather than their direction.

    \item \textit{Number of potential crossings} \cite{brinton1996dynamic}: Number of pairs of aircraft such that, when their trajectories are projected $10$ minutes in the future using their current heading and ground speed, at some moment the two trajectories fall below $10$ NM horizontally. Note that only the horizontal dimension is checked in order to replicate the first-glance evaluation of the ATCO on their screen, i.e. a two-dimensional representation of the sector and the traffic therein \cite{wang2016study}.

    \item  \textit{Number of potential crossings with convergence angle $> 20^{\circ}$} \cite{gianazza2006selection}: Metric computed using the same methodology as the \textit{Number of potential crossings}, but considering only aircraft pairs with a convergence angle greater than $20^{\circ}$. These crossings are generally regarded as more challenging for ATCOs to manage \cite{gianazza2006selection}.

        \item \textit{Number of potential conflicts} \cite{laudeman1998dynamic}: Using the 10-minute projected trajectories used for the \textit{Number of potential crossings} metrics, the number of pairs that generate a potential conflict is counted. A potential conflict corresponds to pairs of aircraft whose projected trajectories fall below $8$ NM horizontally, as defined by the static threshold in Ref.~\cite{gaume2025extracting}; and $900$ ft vertically, to satisfy the vertical condition of a Loss of Separation.

    \item  \textit{Number of potential conflicts with convergence angle $> 20^{\circ}$} \cite{laudeman1998dynamic, gianazza2006selection}: The number of events counted in \textit{Number of potential conflicts} that further satisfy that the conflicting aircraft have a convergence angle greater than $20^{\circ}$. This restriction is again motivated by the idea that those conflicts are generally regarded as more challenging for ATCOs to manage \cite{gianazza2006selection}.

    \item \textit{Number of aircraft with absolute ground speed change $> 10$ kt in $2$ min.} \cite{laudeman1998dynamic}: Number of aircraft changing their ground speed more than $10$ knots, in absolute value, in $2$ minutes.

    \item \textit{Number of aircraft turning} \cite{laudeman1998dynamic}: Number of aircraft that are performing a heading change. This measure reflects the idea that aircraft changing their heading are more complex to manage, as argued in Ref.~\cite{laudeman1998dynamic}. Note that, in the implementation used in this work, the condition is checked at evaluation time, rather than against a future status.

    \item \textit{Number of aircraft climbing} \cite{cardosi1995human}: Number of aircraft that are changing their flight level, either climbing or descending.

    \item \textit{Number of aircraft accelerating} \cite{laudeman1998dynamic}: Number of aircraft that are changing their ground speed, following the same idea behind the metric \textit{Number of aircraft turning}.

    \item \textit{Accumulated sector occupancy} \cite{hurst1978objective}: Total number of aircraft traversing the region under analysis in the $40$-minute window centred on the evaluation time, specifically from $20$ minutes before to $20$ minutes after it.

    \item \textit{Past accumulated sector occupancy} \cite{hurst1978objective}: Total number of aircraft traversing the region under analysis in the 20-minute window before the evaluation time.

    \item \textit{Occupancy variation} \cite{chaboud2000air}: Using the 40 min. time window of \textit{Accumulated sector occupancy}, the instantaneous occupation of the region under analysis is recorded every minute. This metric is then calculated as the standard deviation of the instantaneous occupancies' distribution.

    \item \textit{Maximum simultaneous occupation} \cite{hurst1978objective}: Maximum registered occupation value among the instantaneous occupancies recorded to calculate the \textit{Occupancy variation} metric.

    \item \textit{Mean time flown in sector} \cite{schaefer2001air, majumdar2002factors}: The average time aircraft, which are in the spatial region under analysis at evaluation time, spend on it, since they entered until they will leave.

    \item \textit{Mean distance flown in sector} \cite{schaefer2001air, majumdar2002factors}: The average distance that aircraft, which are in the spatial region under analysis at evaluation time, fly on it, since they entered until they will leave.

    \item \textit{Number of entering aircraft} \cite{majumdar2002factors}: The number of aircraft, from those within the spatial region under analysis at evaluation time, that entered it in the previous 5 minutes.
    
    \item \textit{Number of exiting aircraft}  \cite{majumdar2002factors}: The number of aircraft, from those within the spatial region under analysis at evaluation time, that will exit it in the next 5 minutes.
    
    \item \textit{Number of aircraft} \cite{brinton1996dynamic, masalonis2003dynamic}: Count of the aircraft in the spatial region under analysis at evaluation time.

    \item \textit{Sector area} \cite{gianazza2006selection}: Area of the polygon associated to the spatial region under analysis.

    \item \textit{Sector ID}: Identifier of the spatial region under analysis. Note this identifier is not \textit{per se} a complexity metric, but is still included in the analysis, as it is helpful to identify potential differences in traffic tendencies between spatial regions - as discussed in Sec. \ref{sec:spat_and_temp_effect}.

\end{itemize}

Note that other metrics were tested, but were finally discarded for different reasons. Firstly, \textit{Number of aircraft descending/climbing $750$ ft. in the next $2$ min.} and \textit{Number of aircraft changing their heading $15^{\circ}$ in the next $2$ min} \cite{laudeman1998dynamic} were discarded as they are proxies of the methodology to perform a deconflicting action, respectively vertically and laterally. Secondly, \textit{Mean time to potential crossing/conflict} \cite{chatterji2001measures} and \textit{Mean convergence angle of potential crossings/conflicts} \cite{brinton1996dynamic} were also discarded because they are not defined when no potential crossings/conflicts take place, and no proper default value was found.

As a final methodological note, it is worth highlighting that the Shannon's entropy, when calculated by plugging the empirical distribution into Shannon's formula, is biased. When finite data are available, and not the true distribution, rare symbols are seldom observed, leading to a more peaked empirical distribution; consequently, the entropy is under-estimated, especially in the case of small sample sizes \cite{schurmann2004bias}. This is a general limitation, as there is no perfect entropy estimator: different estimators will perform differently on different data sets, and it is generally impossible to minimise both the variance and the bias \cite{bonachela2008entropy}. While alternatives have been proposed in the literature \cite{feutrill2021review}, for the sake of simplicity, only the plug-in estimator with no bias correction has here been considered.

\section{Machine learning models}
\label{app:class_models}

For the sake of completeness, the four machine learning models used in this work are described below:

\begin{itemize}
    \item Decision Tree (DT). Algorithm that classifies instances by recursively partitioning them according to the input features \cite{maimon2014data, costa2023recent}. It builds a tree-like structure, in which each internal node represents a decision point, and each leaf node holds a class label. In more detail, the root node contains the whole dataset, and its instances are divided based on the attribute that gives the highest information gain (or a similar criterion); this procedure is then repeated until a stopping condition is satisfied. Its main advantages are a low computational cost and high interpretability; nonetheless, in the presence of many features, it is prone to overfitting.

    \item Random Forest (RF). The main idea behind a RF is to construct a group of base learners, usually DT, and aggregate their predictions, so the final outcome is more trustworthy that those given by the individual learners \cite{breiman2001random, biau2016random}. One pivotal feature is the concept of bootstrap aggregation, or bagging. According to it, each DT trains on a subset of the original training dataset, selected via random sampling with replacement. This makes it possible to introduce a variability in each tree, hence allowing them to work with different sections of the data. RF has a smaller risk of overfitting and higher accuracy compared to DT, but also a lower interpretability.

    \item Gradient Boosting (GB). Similarly to the RF model, GB combines multiple base learners, normally DTs. The principal difference is that these learners run sequentially on the data, and not in parallel; in other words, each new model corrects the mistakes made in the previous steps \cite{natekin2013gradient, bentejac2021comparative}. In many situations GB outperforms RF, although at a higher computational cost.

    \item Support Vector Machine (SVM). SVMs are classifiers that perform predictions hinged on decision boundaries in the input space generated by the features, or, more generally, in a transformed feature space \cite{wang2005support, steinwart2008support}. In other words, linear SVMs define a decision boundary to classify, i.e. a line or a hyperplane, which splits data points belonging to different classes; a new instance is then classified as a function of where it is situated with respect to that boundary. With the goal of coping with non-linear relationships, a generalisation is obtained by projecting the instances into a higher-dimensional space through a kernel function, where they are linearly separable.

\end{itemize}

\section{Best-performing models' hyperparameters}
\label{app:all_hyperparameters}

Tab. \ref{tab:optuna_hyperparameters_all} presents the best-performing hyperparameters associated to all the tested ML models and spatial divisions, i.e. the ones used to obtain the accuracies presented in Tab.~\ref{tab:optuna_accuracies_all}.

\begin{table}[!tb]
    \footnotesize
    \centering
    \begin{tabular}{|l|l|c|c|c|c|}
    \hline
    \multicolumn{2}{|c|}{}& Elementary  & Sectors & Area & Swedish  \\
    \multicolumn{2}{|c|}{} & blocks & & Control & FIR\\
    \multicolumn{2}{|c|}{} & & & Centres & \\\hline

        Gradient & learning\_rate $\left[0.01, 1\right]$&$0.015$ & $0.018$ &  $0.026$ & $0.343$ \\
        Boosting & n\_estimators $\left[100,  6400\right]$& $274$ & $282$ & $734$ & $1026$\\
         & min\_samples\_split $\left[2, 200\right]$ & $33$ & $19$ & $22$& $6$\\
         & min\_samples\_leaf $\left[2, 50\right]$ & $4$ & $41$ & $42$ & $17$\\
         & min\_weight\_fraction\_leaf $\left[0.0,  0.5\right]$& $0.0$ & $0.0$& $0.1$ & $0.0$\\
         & max\_depth $\left[3, 320\right]$ & $8$& $31$& $28$ & $43$\\
         & max\_leaf\_nodes $\left[100, 3200 \right]$& $427$ & $528$& $525$ & $741$ \\\hline
         Random & n\_estimators $\left[50,  6400\right]$ & 609 & 54 & 109 & 73\\
         Forest & min\_samples\_leaf $\left[2, 50\right]$ & 3 & 2 & 3 & 2\\
         & max\_leaf\_nodes $\left[3, 320\right]$& 317 & 17 & 34 & 15 \\
         & max\_depth $\left[100, 3200\right]$& 309 & 530 & 439 & 129 \\\hline
         
         Decision  & criterion [``gini'', ``entropy'', ``log\_loss''] & entropy & log\_loss & log\_loss& log\_loss \\
         Tree & min\_samples\_split $\left[2, 200\right]$ & $95$ & $34$& $150$ & $3$\\
         & min\_samples\_leaf $\left[2, 50\right]$& $4$& $32$ & $32$ & $2$\\
         & max\_depth $\left[3, 320\right]$ & $29$& $215$ & $7$ & $33$\\
         & max\_features $\left[2, 3200\right]$ & $1850$ &  $24$ & $1069$ & $21$ \\\hline 
         SVC & c\_value $\left[0.001, 100\right]$ &  $68.58$ & $0.067$ & $36.92$ &  $0.052$\\
         & kernel [``rbf'', ``linear'', ``poly'', ``sigmoid''] & rbf & rbf & rbf & rbf \\
         & degree $\left[2, 5\right]$& $3$ & $4$ & $5$ & $5$\\
         & gamma [``scale'', ``auto''] & scale & scale & auto & auto\\
         & max\_iter  $\left[5, 50000\right]$& $26,372$ & $34,789$ & $801$  & $447$ \\\hline 
         
         Linear &  loss [``hinge'', ``squared\_hinge''] & hinge & hinge & hinge & squared\_hinge\\
         SVC & c\_value $\left[0.001, 100\right]$ & $0.001$ & $0.009$ & $5.975$ & $0.402$ \\
         & intercept\_scaling $\left[1, 100\right]$ & $6.32$& $27.29$& $6.04$ & $54.99$  \\
         & max\_iter $\left[1000, 50000\right]$& $3,805$& $1,077$ & $1, 001$ & $21,965$ \\\hline
         
    \end{tabular}
    \caption{ Hyperparameters for the highest accuracy models, as identified by \textit{Optuna}, for each tested ML model and each spatial scale. The range of values allowed in the optimisation process are reported alongside each hyperparameter name.}
    \label{tab:optuna_hyperparameters_all}
\end{table}

\section{Analysis of elementary blocks}
\label{app:ind_sec_vol}

This section delves deeper in the result presented in the left panel of Fig.~\ref{fig:area_score_evolution} of Sec.~\ref{sec:spat_and_temp_effect}, specifically regarding the elementary blocks. In that figure it was observed that the accuracy of the model and its evolution, when using only the airspace-related features, was qualitatively the same as when all complexity metrics were included as features. We here analyse the difference in the airspace structure metrics' distributions between deconflictions and conflict-free events, in order to shed light on the causes behind such result.

\begin{figure}[!tb]
     \centering
     \includegraphics[width=\linewidth]{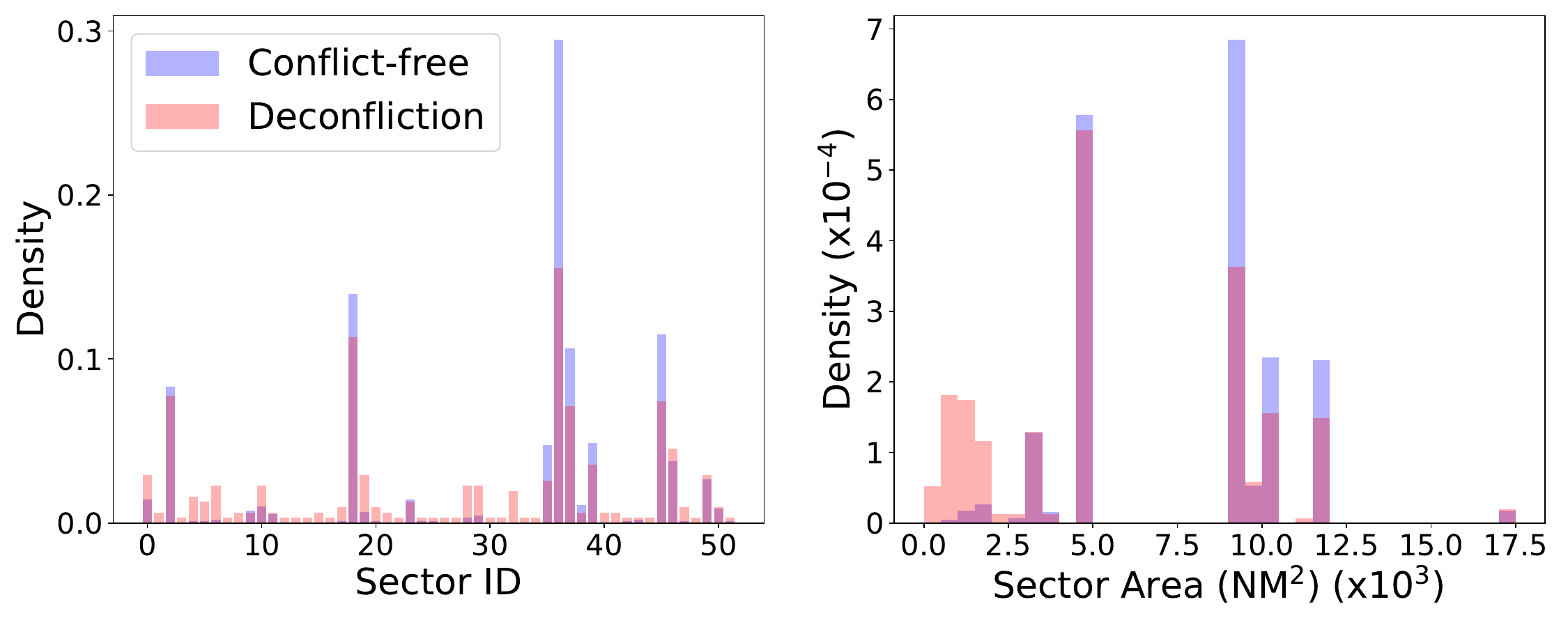}
     \caption{Comparison of the density of deconflictions and conflict-free events at the elementary block level. The left panel depicts the density of both types of events on each elementary block, sorted by the corresponding ID. The right panel presents the density of events as a function of the area of the corresponding elementary block. Results correspond to the full data set at clearance time. }
     \label{fig:volumes_sectors_ids_and_areas_distributions}
 \end{figure}

The left panel of Fig.~\ref{fig:volumes_sectors_ids_and_areas_distributions} reports the density of deconfliction and conflict-free events, where each bar is a unique elementary block - as defined by the \textit{Sector ID}. 
While conflict-free events seem to be clustered in a reduced number of blocks, deconflictions are more dispersed. In other words, it can be appreciated that many elementary blocks have experienced a handful of deconflictions, but no conflict-free events.
Similar results are reported in the right panel of Fig.~\ref{fig:volumes_sectors_ids_and_areas_distributions}, here as a function of the area of each block. They further show that the elementary blocks in which the imbalance between both groups is observed correspond to the smallest ones.

These results are easily explained by considering the data set construction methodology, and specifically how valid events are selected for balancing the two classes (see Sec.~\ref{sec:class_prob_def}). 
In elementary blocks of small area, most conflict-free instances have no traffic (or only one aircraft). They thus cannot be added to the data set, and deconfliction events have no counterpart there. In order to maintain a balance, conflict-free events have to be more frequently drawn from large-area elementary blocks, which consistently have two or more aircraft. 
The final result is a bias, with deconfliction events appearing more frequently in elementary blocks of small area, and conflict-free scenarios in large ones. The machine learning model exploits the block size/identity to separate classes, and hence yields good classification results even when only including these two features.

\section{Metrics' normalisation}
\label{app:norm_metrics}

Sec.~\ref{sec:stab_of_results} presented results relative to the stability of the analysis, with one of them requiring the normalisation of the complexity metrics used as features in the ML models, to remove sectors' geometry or traffic effects (see left panel of Fig.~\ref{fig:different_normalizations_score_evolution}). Tab.~\ref{tab:metrics_normalization} presents, when applicable, the exact normalisation methodology followed for each complexity metric.

\begin{table}[!tb]
\centering
\begin{tabular}{l | l}
\hline
        Metric & Normalised with \\
        & (Divided by) \\\hline
Mean horizontal distance & Maximum horizontal distance  \\
& achievable in the sector \\
Mean vertical distance & Altitude range of the sector\\
Pairwise horizontal  & Entropy of the most heterogenous   \\
distances entropy & situation  achievable in the sector  \\
Number of aircraft close & Number of aircraft \\
 to sector border &  \\
Mean traffic convergence/divergence & None \\
Interacting pairs & Number of pairs of aircraft\\
Number of different flight levels in use& None\\
Flight levels occupancy entropy &  None\\
Heading angle coherence & None \\
Mean ground speed & None\\
Ground speeds variance & None \\
Mean vertical speed & None \\
Number of potential crossings &  Number of pairs of aircraft \\
Number of potential crossings & Number of pairs of aircraft\\
with convergence angle $> 20^{\circ}$ & \\ 
Number of potential conflicts & Number of pairs of aircraft \\
Number of potential conflicts & Number of pairs of aircraft\\
with convergence angle $> 20^{\circ}$ & \\ 
Number of aircraft with absolute & Number of aircraft\\
ground speed change $>$ 10 kt in 2 min &  \\
Number of aircraft climbing & Number of aircraft \\
Number of aircraft turning  & Number of aircraft\\
Number of accelerating planes & Number of aircraft \\
Accumulated sector occupancy & Area of the sector \\
Past accumulated sector occupancy & Area of the sector \\
Occupancy variation & None \\
Maximum simultaneous occupation & Area of the sector \\
Mean time flown in sector & Area of the sector\\
Mean distance flown in sector & Area of the sector  \\
Number of entering aircraft & Number of aircraft \\
Number of exiting aircraft & Number of aircraft\\
Number of aircraft& Area of the sector\\
\hline
\end{tabular}

\caption{Normalisation methodology followed for each studied complexity metric.}
\label{tab:metrics_normalization}
\end{table}

\section{Correlation between metrics}
\label{app:metrics_correlation}

\begin{figure}[!tb]
    \centering
    \includegraphics[width=\linewidth]{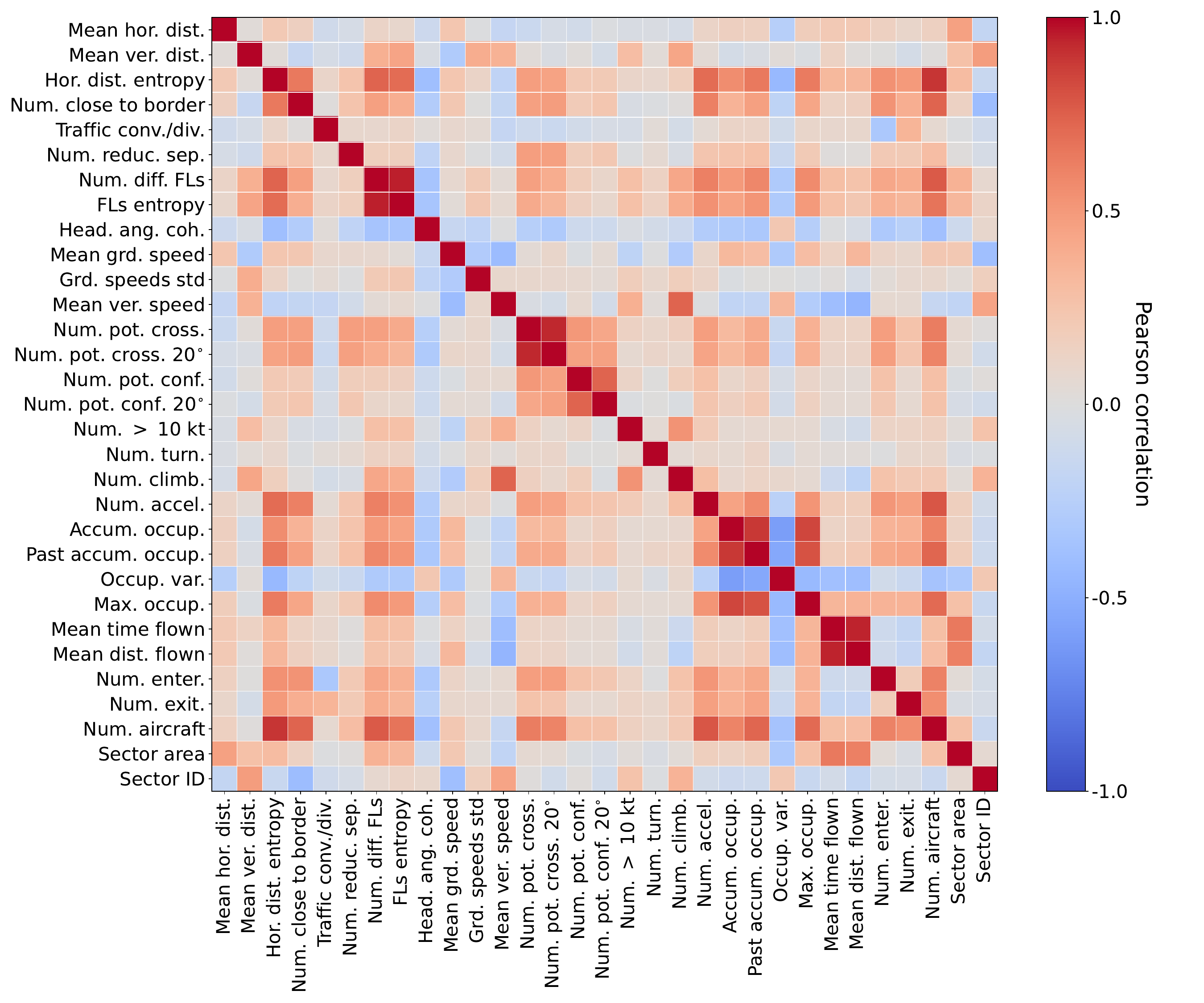}
    \caption{Pearson's linear correlation between all pairs of complexity metrics, for the subset of deconflicting events. Note that, for the sake of compactness, the names of the metrics have been shortened. To avoid confusions, metrics are reported in the same order as in Tab.~\ref{tab:compl_metrics}.}
    \label{fig:pearson_correlation_severe}
\end{figure}

\begin{figure}[!tb]
    \centering
    \includegraphics[width=\linewidth]{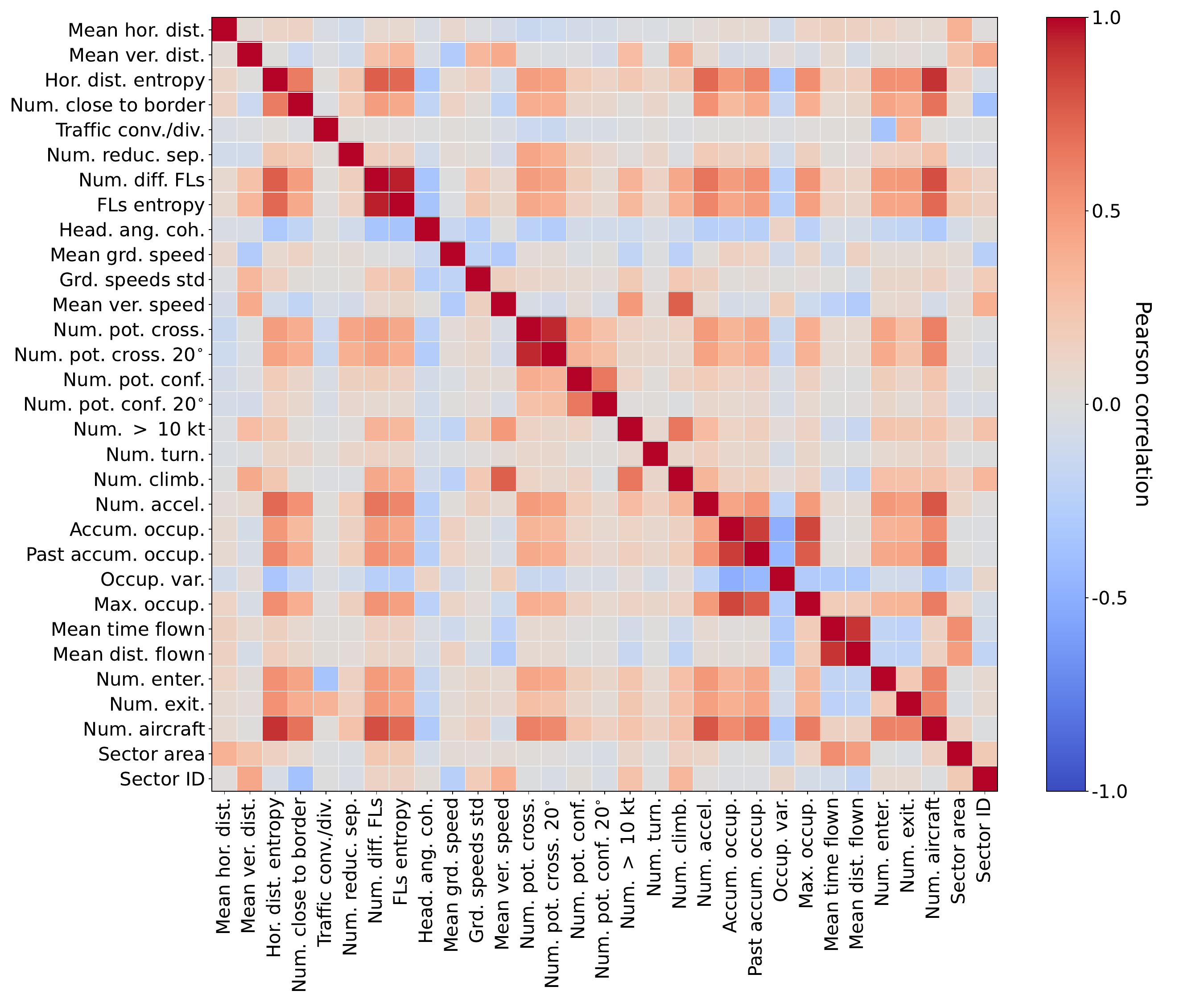}
    \caption{Pearson's linear correlation between all pairs of complexity metrics, for the subset of conflict-free events. Note that, for the sake of compactness, the names of the metrics have been shortened. To avoid confusions, metrics are reported in the same order as in Tab.~\ref{tab:compl_metrics}.}
    \label{fig:pearson_correlation_non_severe}
\end{figure}

This appendix presents the results of the correlation between the complexity metrics that have been used throughout this work. Specifically, Fig.~\ref{fig:pearson_correlation_severe} shows the Pearson's linear correlation matrix for the deconflicting events, while Fig.~\ref{fig:pearson_correlation_non_severe} the one corresponding to conflict-free scenarios. Without considering the magnitude, for both types of events, the correlation between metrics is overall positive. Only \textit{Heading angle coherence} and \textit{Occupancy variance} have systematic negative correlations.
The largest observed correlations occur mainly between metrics which share concepts in their definitions. This is the case, for instance, of \textit{Number of different flight levels in use} and \textit{Flight levels occupancy entropy}; notably, such high correlation implies that little additional information is gained by considering both of them, and specifically by considering how many aircraft beyond the first are in each given flight level. 
Other examples include the family of metrics that consider the airspace occupation at different accumulated temporal scales, i.e. \textit{Accumulated sector occupancy}, \textit{Past accumulated sector occupancy}, \textit{Occupancy variation}, and \textit{Maximum simultaneous occupation}.
Note, nonetheless, that the correlation between these metrics and \textit{Number of aircraft} is not as large, suggesting that the status of the airspace changes as a function of the time scale in which it is analysed. 
The high correlation between \textit{Mean time flown in sector} and \textit{Mean distance flown in sector} suggests that a narrow range of ground speeds is observed among the traffic.
The correlations between \textit{Number of potential crossings} and \textit{Number of potential conflicts}, and their $20^{\circ}$-filtered counterparts, show that each pair of metrics vary almost proportionally.
Finally, \textit{Number of aircraft} highly correlates with many other metrics (e.g. \textit{Number of aircraft accelerating}, \textit{Number of different flight levels in use}). These correlations are to be expected, as these  metrics have in common that their value is highly influenced by the traffic volume. Nonetheless, the analysis presented in Sec.~\ref{sec:stab_of_results} shows that, despite such large correlations, the traffic volume is not exclusively driving the classification performance, i.e. that there is extra information in those complexity metrics.

\begin{figure}[!tb]
    \centering
    \includegraphics[width=\linewidth]{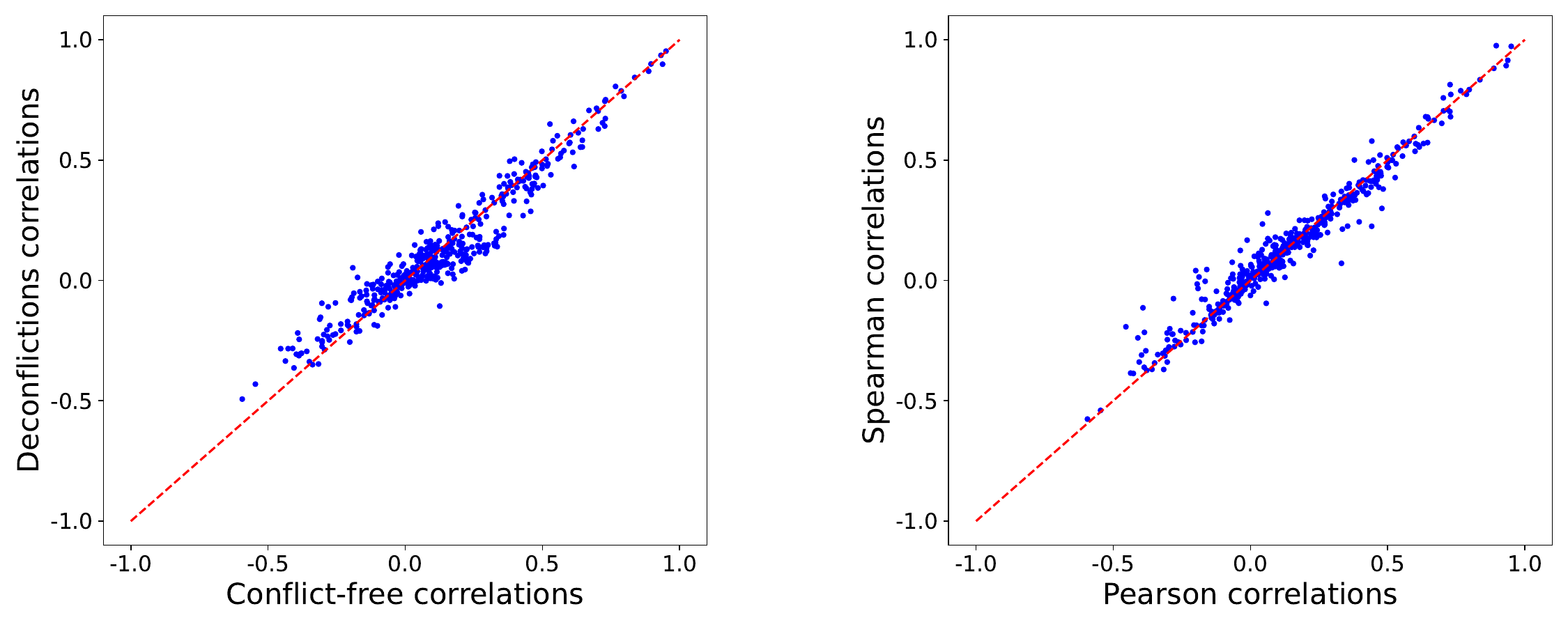}
    \caption{Comparison of correlations. The left panel shows a scatter plot comparing the values of the Pearson's linear correlation for deconfliction events, as a function of what obtained for conflict-free events. The right panel shows, for the deconfliction events, a scatter plot of the Spearman's rank correlation values as a function of the Pearson's ones. The red dashed lines corresponds to the main diagonal of the plots, $y=x$.}
    \label{fig:correlations_comparisons}
\end{figure}

For the sake of completeness, Fig.~\ref{fig:correlations_comparisons} presents additional results regarding the correlation between complexity metrics. On the one hand, the left panel depicts a scatter plot, comparing the Pearson's linear correlations obtained, for each pair of complexity metrics, in deconflicting and conflict-free events. On the other hand, the right panel compares, for each pair of complexity metrics and for deconflicting events, the Pearson's and Spearman's correlations. A clear agreement with the $y=x$ reference diagonal is present in both. This indicates, in the former case, that the correlation structure is event-independent, and the metrics behave consistently across both classes; and, in the latter case, that the correlations are mainly linear.

\section{Metrics' distributions difference}
\label{app:met_dist_diff}

Tab.~\ref{tab:significant_metrics_distributions_differences} reports the Wasserstein distance corresponding to all complexity metrics, for the analysis of the metrics' distributions presented in Sec.~\ref{sec:metrics_distributions_comparisons}.

\begin{table}[!tb]
\centering
\resizebox{\linewidth}{!}{
\begin{tabular}{l c c c | c}
\hline
        Metric & Deconflictions  & Clearance time & Tendency & Clearance time\\
                 & vs conflict free & vs 8 min.  & & vs 8 min.\\

        & at clearance & before.  & & before.\\
                & time (W) & Deconf. (W) & & Conf. free (W)\\
\hline
Mean horizontal distance & $0.055$ & $0.062$ & - & $0.034$ \\
Mean vertical distance & $0.057$ &  $0.060$ & - & $0.053$ \\
Pairwise horizontal  & $0.266$ & $0.244$ & $\uparrow$ & $0.060$ \\
distances entropy & & & & \\
Number of aircraft close & $0.311$ & $0.297$ & $\uparrow$ & $0.034$ \\
 to sector border & & & & \\
Mean traffic convergence/divergence & $0.138$ & $0.058$ & - & $0.065$ \\
Number of pairs with & $0.254$ & $0.294$ & $\uparrow$ & $0.011$\\
reduced separation & & & & \\
Number of different flight levels in use& $0.167$& $0.137$ & - & $0.069$\\
Flight levels occupancy entropy &  $0.136$ & $0.113$ & - & $0.064$\\
Heading angle coherence & $0.102$ & $0.132$ & - & $0.020$ \\
Mean ground speed & $0.183$ & $0.114$ & - &  $0.050$\\
Ground speeds variance & $0.136$ & $0.071$ & - & $0.020$ \\
Mean vertical speed & $0.064$ & $0.078$ & - & $0.043$ \\
Number of potential crossings &  $0.339$ & $0.336$ & $\uparrow$ & $0.021$ \\
Number of potential crossings & $0.283$ & $0.305$ & $\uparrow$ & $0.023$\\
with convergence angle $> 20^{\circ}$ & & & & \\ 
Number of potential conflicts &  $0.353$ & $0.318$ & $\uparrow$ & $0.063$ \\
Number of potential conflicts & $0.283$ & $0.254$ & $\uparrow$ & $0.037$\\
with convergence angle $> 20^{\circ}$ & & & & \\ 
Number of aircraft with absolute & $0.387$ & $0.309$ & $\uparrow$ & $0.024$ \\
ground speed change $>$ 10 kt in 2 min & & & & \\
Number of aircraft climbing & $0.075$ & $0.037$ & - & $0.026$ \\
Number of aircraft turning  & $0.032$ & $0.050$ & - & $0.024$ \\
Number of accelerating planes & $0.215$ & $0.152$ & $\uparrow$ & $0.038$ \\
Accumulated sector occupancy & $0.061$ & $0.151$ & - & $0.096$\\
Past accumulated sector occupancy & $0.060$ & $0.069$ & - & $0.069$ \\
Occupancy variation & $0.200$ & $0.168$ & $\uparrow$ & $0.091$\\
Maximum simultaneous occupation & $0.054$ & $0.170$ & - & $0.088$ \\
Mean time flown in sector & $0.251$ & $0.256$ & $\downarrow$ &  $0.048$\\
Mean distance flown in sector & $0.273$ & $0.245$ & $\downarrow$ &  $0.064$  \\
Number of entering aircraft & $0.352$ & $0.209$ & $\uparrow$ & $0.034$\\
Number of exiting aircraft & $0.258$ & $0.355$ & $\uparrow$ & $0.052$\\
Number of aircraft& $0.289$ & $0.270$ & $\uparrow$ & $0.064$\\
Sector Area & $0.128$ & $0.094$ & -  & $0.051$\\
Sector ID & $0.098$ & $0.050$ & - & $0.026$ \\\hline
\hline
\end{tabular}
}
\caption{Wasserstein distance values for all complexity metrics, in the three comparisons defined in the main text - see Tab. \ref{tab:significant_metrics_distributions_differences}. The tendency in the fifth column is only reported for those metrics with $W\geq 0.2$. Results correspond to the full data set with non-normalised metrics. See Sec.~\ref{sec:metrics_distributions_comparisons} for a detailed explanation of each column's content.
}
\label{tab:metrics_distributions_differences}
\end{table}

\newpage
\clearpage

\printbibliography

\end{document}